\documentclass[useAMS,usenatbib]{mn2e}

\usepackage{graphicx}
\usepackage{wasysym}
\usepackage{hyperref}
\usepackage{subfig}
\usepackage{color}

\newcommand{\bo}[1]{\mathbf{#1}}

\newcommand{\beq}{\begin{equation}}
\newcommand{\eeq}{\end{equation}}
\newcommand{\Mpc}{\,h^{-1}\,\mathrm{Mpc}}
\newcommand{\iMpc}{\,h\,\mathrm{Mpc}^{-1}}
\newcommand{\Msun}{\,h^{-1}\,\mathrm{M_\odot}}
\newcommand{\kpc}{\,h^{-1}\,\mathrm{kpc}}
\newcommand{\lcdm}{$\Lambda$CDM\ }
\newcommand{\tcdm}{$\tau$CDM\ }
\newcommand{\sinc}{\mathrm{sinc}\,}

\newcommand{\eg}{\emph{e.g.}\ }
\newcommand{\ie}{\emph{i.e.}\ }
\newcommand{\etc}{\emph{etc.}\ }
\newcommand{\mrm}[1]{\mathrm{#1}}
\newcommand{\sform}[2]{{#1}\times 10^{#2}}
\newcommand{\average}[1]{\langle{#1}\rangle}

\def\m@th{\mathsurround=0pt }
\def\eqalign#1{\null\,\vcenter{\openup1\jot \m@th
 \ialign{\strut\hfil$\displaystyle{##}$&$\displaystyle{{}##}$\hfil
 \crcr#1\crcr}}\,}

\title[Remapping dark matter halo catalogues between cosmological simulations]{Remapping dark matter halo catalogues between cosmological simulations}
\author[A. J. Mead and J. A. Peacock]{A. J. Mead$^{1}$\thanks{E-mail: am@roe.ac.uk} and J. A. Peacock$^{1}$\\
$^{1}$Institute for Astronomy, University of Edinburgh, Royal Observatory, Blackford Hill, Edinburgh EH9 3HJ\\}

\begin{document}

\date{Accepted 2014 February 20.  Received 2014 February 20; in original form 2013 August 23}
\pagerange{\pageref{firstpage}--\pageref{lastpage}} 
\pubyear{2014}

\maketitle

\label{firstpage}

\begin{abstract}
We present and test a method for modifying the catalogue of dark
matter haloes produced from a given cosmological simulation, so that
it resembles the result of a
simulation with an entirely different set of parameters. This extends
the method of \cite{Angulo2010}, which rescales the full particle
distribution from a simulation. Working directly with the
halo catalogue offers an advantage in speed, and also allows
modifications of the internal
structure of the haloes to account for nonlinear differences between
cosmologies. Our method can be used directly on a halo catalogue in a self contained manner
without any additional information about the overall density field;
although the large-scale displacement field is required by the method,
this can be inferred from the halo catalogue alone.
We show proof of concept of our method by rescaling a
matter-only simulation with no baryon acoustic oscillation (BAO)
features to a more standard \lcdm model containing a cosmological
constant and a BAO signal. In conjunction with the halo occupation
approach, this method provides a basis for the rapid generation of mock
galaxy samples spanning a wide range of cosmological parameters.
\end{abstract}


\section{Introduction}

Ever since the first measurements of the accelerated expansion of the
cosmos from supernova data (\citealt{Schmidt1998};
\citealt{Perlmutter1999}), the origin of the acceleration
has been a dominant, open question in cosmology. 
Either space is filled with a nearly homogeneous substance,
dark energy, or we are witnessing a breakdown of Einstein's
relativistic gravity on cosmological scales. Exploring these
issues, and looking for deviations from a pure cosmological
constant, requires us to measure the global expansion history
precisely, together with the rate of growth of density fluctuations.
The expansion history involves standard candles and standard rulers,
especially the signal of BAO (Baryon Acoustic Oscillations) in
the galaxy distribution (\eg \citealt{Anderson2013}); the
growth rate can be probed directly by
gravitational lensing (\eg CFHTLens: \citealt{Heymans2012}) or via
redshift-space distortions in galaxy clustering
(\eg \citealt{Samushia2013}; \citealt{vipers2013}).

The extraction of this fundamental cosmological information
increasingly requires a major input from cosmological
$N$-body simulations, for two reasons. The statistical quantities
to be measured from the data tend to have complicated correlations,
and the only practical way of computing the required covariance
matrix is by averaging over an ensemble of mock datasets.
More seriously, an analytical understanding of the development
of cosmological structure is restricted to large-scale linear
fluctuations, whereas the measurements are inevitably affected
by small-scale nonlinearities to some extent.
The mildly nonlinear regime can be explored with perturbation theory
(\eg \citealt{Bernardeau2002}) but this fails on smaller scales. If
nonlinear information is to be exploited, it is necessary to run
simulations for different sets of cosmological
parameters, to measure the matter distribution and derive halo catalogues.
Mock galaxy samples can then either be
generated using semi-analytic methods (\eg \citealt{Baugh2006}) or
from halo occupation distribution models (\citealt{Seljak2000};
\citealt{Peacock2000}; \citealt{Zheng2005}).

However, it is computationally prohibitive to run simulations of large
enough volumes at a high enough resolution in order to cover the
current cosmological parameter space, which has now grown to encompass
neutrinos (masses and numbers of species); warm
dark matter; plus complex dark energy models and modified gravity
theories.
We therefore need some way of spanning this range of cosmologies without
having to run a simulation for each particular set of parameters. This
idea was investigated by \citeauthor{Angulo2010} (\citeyear{Angulo2010}; 
hereafter AW10), who showed that it was possible to rescale an
$N$-body particle distribution 
in order to approximate the results of a simulation with a different
set of cosmological parameters. Their algorithm consisted of two
steps: (i) reinterpreting the length and time units in the original
simulation so that the halo mass function was as close as possible to
that which would be measured in the new cosmology (ii) modifying
individual particle positions so as to reproduce the expected linear
clustering in the new cosmology.

AW10 showed that their method successfully reproduced the
statistics of the target cosmology at the level of the matter power
spectrum and halo mass function. AW10 has been applied by
\cite{Guo2013} to look at theoretical differences in galaxy formation
between WMAP1 and WMAP7 cosmologies and by
\cite{Simha2013}, who looked at measuring cosmological parameters by
comparing the galaxy two-point correlation function of SDSS with that
computed from galaxy catalogues that were rescaled using the AW10
method.

Despite the success of the AW10 algorithm, it has some
disadvantages. Firstly, the algorithm is applied to large particle datasets that
can be difficult to communicate;
often it is only halo catalogues that are made
publicly available by large collaborative simulation groups (\eg the
DEUSS simulations of \citealt{DEUSS}). Secondly, the algorithm uses the
displacement field that was employed to generate the initial conditions;
again this may not be publicly available. Finally, the algorithm reproduces the
linear clustering in the target simulation, but does not
reproduce the deeply nonlinear clustering, which can be considered to be
associated with correlations within individual haloes.  In this paper
we develop and test an extension to the AW10 algorithm designed to
deal with these issues.

Our method rapidly converts a halo catalogue from a given simulation into 
one that is characteristic of a different
cosmology. Other methods for the fast generation of halo catalogues
have been developed in the literature: \cite{Monaco2002} developed an
algorithm called \texttt{PINOCCHIO}, which uses a combination of perturbation
theory and an ellipsoidal halo collapse model to generate catalogues.
 \cite{Manera2013} produced mock 
catalogues for the Baryon Oscillation Spectroscopic
Survey (BOSS) using second order Lagrangian perturbation theory (2LPT)
on a particle distribution and then collecting mass from the evolved
field into haloes; this approach is called
\texttt{PTHaloes}. \cite{Tassev2013} use an approach called
\texttt{COLA} (COmoving Lagrangian Acceleration), which involves
a coordinate transform based on 2LPT, followed by a
particle mesh (PM) gravity solver with coarse time-stepping, which is
able to yield halo statistics rapidly. Nevertheless, all these methods are
approximate in their treatment of nonlinearities, and an attractive feature
of AW10 is that it is based on a fully nonlinear simulation. A reduced version of the AW10
method has been applied to halo catalogues by \cite{Ruiz2011}, in which
the authors scaled a halo catalogue in time and length
units but did not apply the final stage of the algorithm, in
which the linear clustering is reproduced by modifying individual halo
positions. 
In this case \cite{Ruiz2011} showed that AW10 works very well on halo
catalogues, but only for simulations of small box sizes ($<50\Mpc$) in which large-scale 
shifts in the displacement field are unimportant and would only manifest
themselves as translations of the entire box. Nevertheless the
authors showed that halo positions and velocities were
recovered with almost no detectable biases and information useful for galaxy formation
modelling, such as merger histories, could also be accurately recovered.

Our extended algorithm consists of the following steps: The length and
time units in the original halo catalogue are rescaled exactly as in
the original AW10 algorithm. We then use the particles or halo
distribution itself to compute the linear displacement field,
from which we modify the particle or halo
positions so that they reproduce the correct large-scale clustering in
the target cosmology. 
\cite{Eisenstein2007} showed how to
recreate the displacement fields via the over-density field in a
simulation by using a reverse of the approximation due to
\cite{Zeldovich1970}. In \cite{Padmanabhan2012} a variant of this
approach was used to improve the sharpness of the BAO
feature in BOSS data. Finally, we modify
the halo internal structure directly -- either by `reconstituting' the
density profiles around haloes so that they have the correct sizes and
internal structure for the target cosmology, or by removing halo particles from
the scaled particle distribution and then `regurgitating' them with
the correct internal structure back into the distribution of non-halo
particles. In this way we are able to create consistent particle
and halo distributions for any desired cosmological model. It is important to
emphasise that we are able to do this in an entirely self-contained manner from only
a pre-existing halo catalogue and without any tuned parameters.

Our paper is set out as follows: In Section \ref{sec:method} we review
the AW10 algorithm and explain our extensions to it. 
In Section \ref{sec:recasting}, we discuss the cosmology dependence 
of the internal structure of haloes.
In Section
\ref{sec:simulations} we describe our simulations and our methods for
generating halo catalogues.
In Section \ref{sec:results} we first show
that our method for computing the displacement field from the halo
positions is reasonable and then show results for the mass functions,
clustering of matter, clustering of haloes and clustering of material
in the interiors of haloes. Finally we sum up in Section
\ref{sec:conclusions}.

\section{Rescaling}
\label{sec:method}

The first part of the AW10 algorithm relabels redshifts and
rescales the box size in the original simulation, so that the
halo mass function becomes as similar as possible to 
the desired target cosmology over the
range of masses probed by haloes in the simulations. Cosmological mass
functions have been shown to be nearly universal in form (\eg
\citealt{Sheth1999}; \citealt{Tinker2008}) and depend on cosmology
almost entirely through the linear variance, defined in equation (\ref{eq:sigma}),
and which in turn depends only on the linear power spectrum. 
Because the CDM power spectrum is continuously curved,
a suitable scaling in redshift and length units can always
make the linear variances as a function of smoothing scale in two
different cosmologies coincide almost perfectly around the nonlinear scale.
In this way, the re-interpreted simulation output should have the
desired halo mass function. This is closely related to the
small-scale nonlinear power spectrum via
the one-halo
term in the halo model (\citealt{Seljak2000}; \citealt{Peacock2000}),
where structure is considered to be made of a distribution of clustered virialized
haloes with a certain internal structure and mass distribution. If the re-interpreted
simulation has the correct mass function then the one-halo term should be approximately
correct. The
two-halo term in the power is essentially the linear clustering of
matter, and this will not be perfectly reproduced by the rescaling.

The second part of the AW10 algorithm therefore aims to correct this latter problem,
using the approximation of \cite{Zeldovich1970} to displace individual
particles so that the linear clustering is exactly matched. As pointed
out in AW10, one of the remaining sources of difference between the
two cosmologies after this scaling will be the different internal halo
structure caused in part by the haloes being concentrated differently
due to collapsing at different redshifts depending upon the background
cosmology, thus altering the one-halo term. We address this by
modifying the internal structure of the haloes directly so that we can
update the structure to that of the new cosmology. We do this either
by equipping catalogued haloes with the correct internal
structure for the new cosmology (a method we call reconstitution) or by finding
halo particles in the scaled particle distribution and replacing these with
a set of particles designed to have the correct internal structure (a method we call regurgitation).

\begin{figure}
\begin{center}
\includegraphics[width=60mm,angle=270]{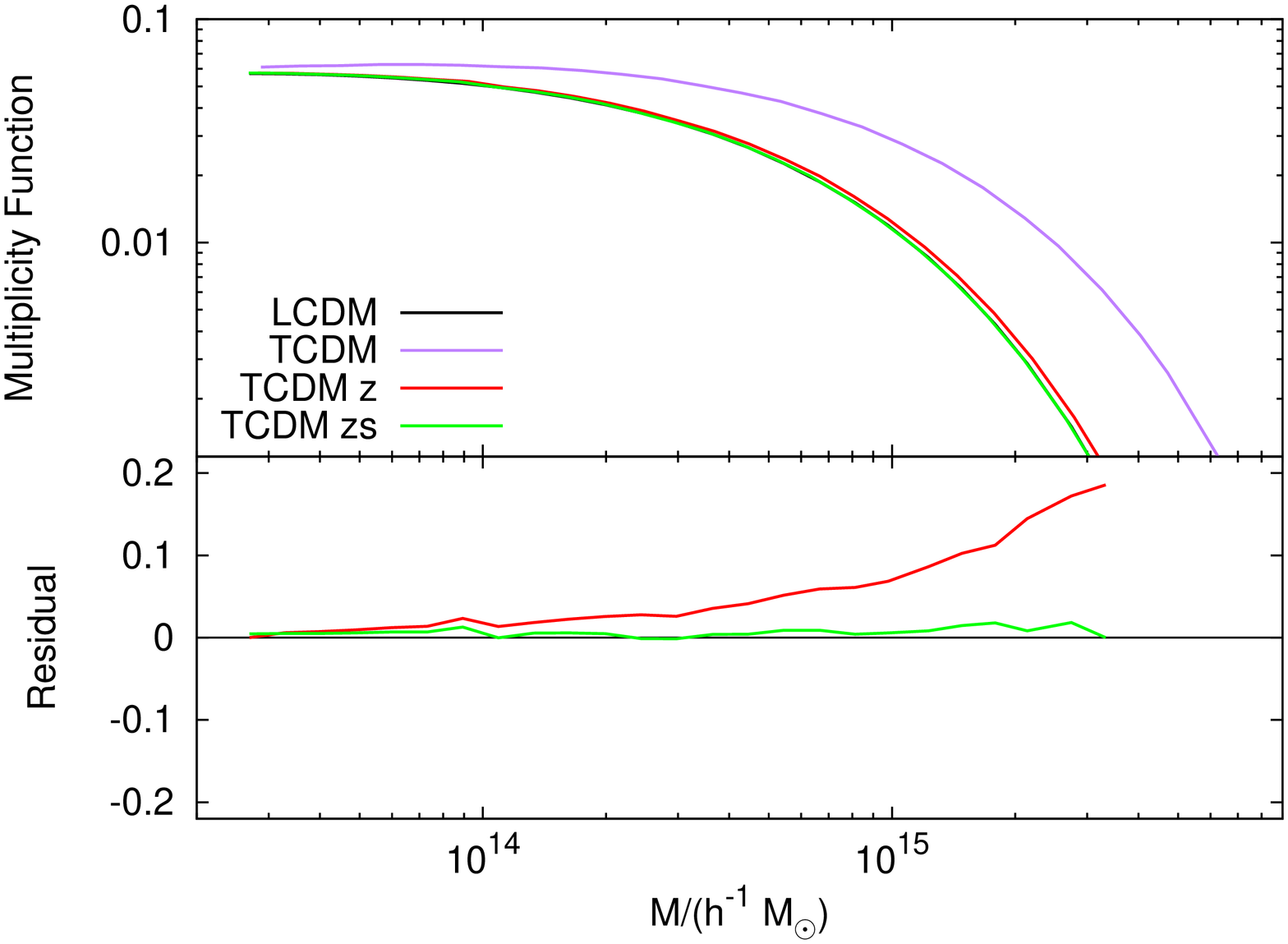}
\includegraphics[width=60mm,angle=270]{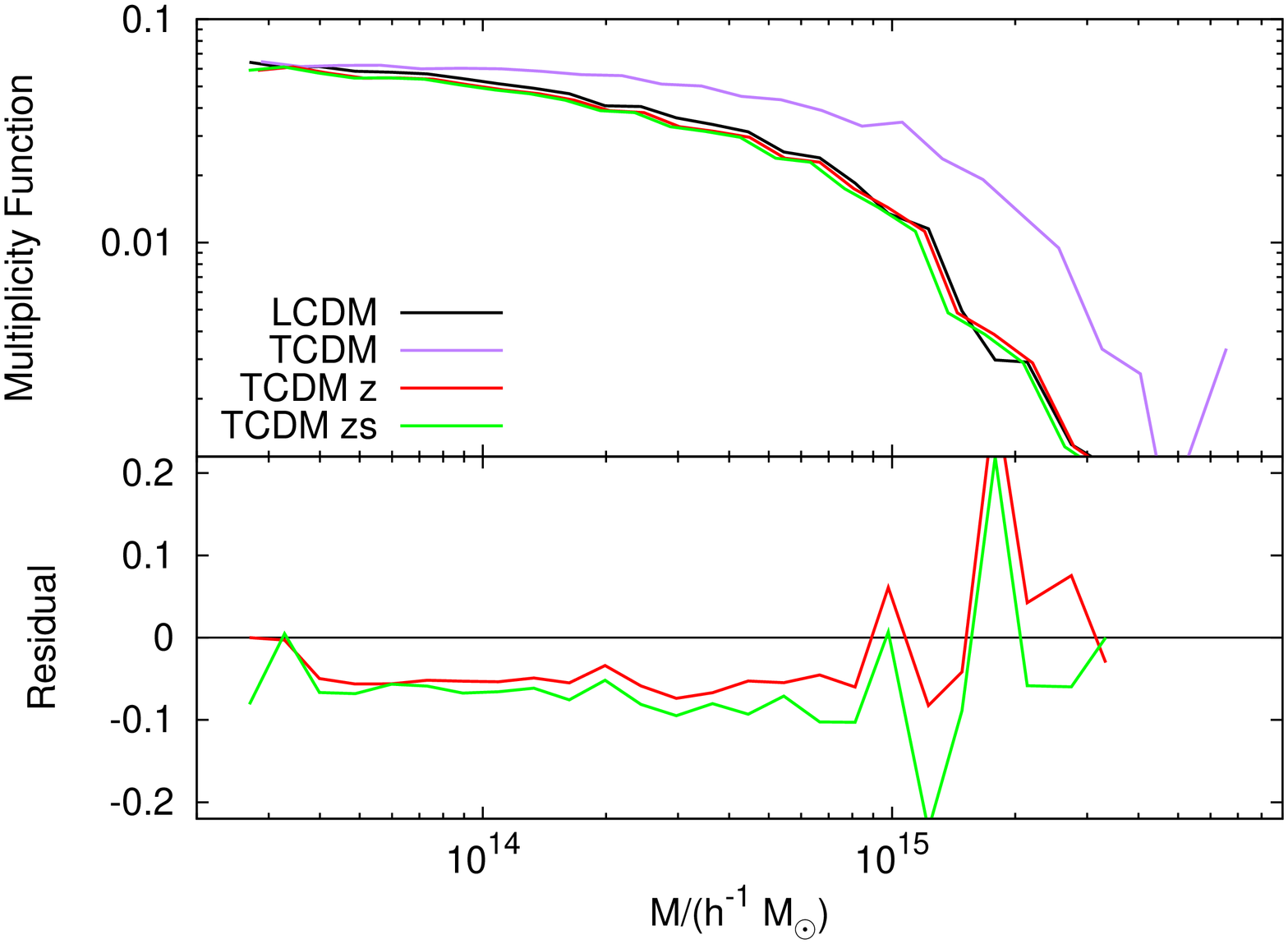}
\end{center}
\caption{Halo mass functions before and after the scaling
  procedure. The top panel shows the theoretical mass function of
  \citet{Sheth1999}, whereas the bottom panel shows measurements from
  our simulations (discussed in Section \ref{sec:simulations}). In
  each panel the mass functions are shown for the target \lcdm
  cosmology (black); the original \tcdm cosmology (purple); the
  effect of relabelling the redshift of \tcdm (red); and the effect of
  then also scaling the simulation box size (green), which
  simultaneously changes individual halo masses. 
  The values of the scaling parameters $z$ and $s$ used to achieve 
  this are given in Table \ref{tab:scaling} and the details of the simulations
  are discussed in section \ref{sec:simulations}.
  We also show the fractional residual between the mass function in the scaled
  simulations and in the target cosmology; this does not
  vanish perfectly for the simulation data, indicating
  that the mass function is not perfectly
  universal at the few per cent level.}
\label{fig:mf}
\end{figure}

\subsection{Matching the mass function}

Throughout this paper quantities in the target cosmology are denoted
with primes and quantities in the original simulation are unprimed.

The AW10 algorithm first chooses a rescaling in length units of the
simulation such that 
\beq
L'=sL
\label{eq:size_scaling}
\eeq
and a rescaling in redshift so that
outputs in the original simulation at redshift $z$ are matched to a
different redshift in the target simulation $z'$. 
Note that we will assume that the box side, $L$, is measured in
comoving units, so that $s$ rescales all comoving lengths.
We also choose units of $\Mpc$ for $L$;
this is not mandatory, but it simplifies some related scalings,
such as that of mass (equation \ref{eq:mass_scaling}). The appropriate
powers of $h$ must then be carried in the units of all quantities,
such as $\Msun$.

For a given $z'$, $s$ and $z$ are
chosen so as to minimize the difference in the halo mass function
between the two cosmologies. To achieve this the RMS difference in the
linear variance in density between the two cosmologies is minimized
over both $s$ and $z$: 
\beq
\delta_{\mrm{rms}}^2(s,z\mid z')=\frac{1}{\ln(R'_2/R'_1)}\int_{R'_1}^{R'_2}
\frac{\mrm{d}R}{R}\left[1-\frac{\sigma(R/s,z)}{\sigma'(R,z')}\right]^2\ ,
\label{eq:minimise}
\eeq 
where $R'_1$ and $R'_2$ are the radial scales,
measured in the target cosmology, corresponding to the least massive
and most massive haloes in the original catalogue. The radial
scale $R$ is given by the radius that would enclose a mass $M$ in a
homogeneous Universe, 
\beq 
M=\frac{4}{3}\pi R^3 \bar{\rho}\ , 
\eeq
where $\bar{\rho}$ is the mean cosmological matter density. Scales in
the two simulations are related by $R'=sR$; this
size rescaling here thus implies a
rescaling of the mass via
\beq
M'=s_{\mrm m} M;
\quad s_{\mrm m} \equiv s^3\frac{\Omega'_{\mrm m}}{\Omega_{\mrm m}}\ ,
\label{eq:mass_scaling}
\eeq 
such that the total mass enclosed in the simulation volume matches 
the cosmological mass after the rescaling has been applied. Again note
that our definition of $M$ includes the units $\Msun$.

The linear variance in over-density can be expressed in Fourier space as 
\beq 
\sigma^2(R,z)=\int_0^{\infty}
\Delta_{\rm{lin}}^2(k,z)T^2(kR)\;\mrm{d}\ln{k}\ ,
\label{eq:sigma}
\eeq
where $T(kR)$ is the Fourier transform of a top-hat of radius $R$ 
which contains a mass $M$ in a homogeneous universe:
\beq
T(kR)=\frac{3}{(kR)^3}(\sin kR-kR\cos kR)\ .
\eeq
$\Delta_{\rm{lin}}^2(k,z)$ is the dimensionless, linear matter power spectrum 
\beq
\Delta_{\rm{lin}}^2(k,z)=4\pi V\left(\frac{k}{2\pi}\right)^3 P(k,z)\ ,
\eeq
giving the variance per unit $\Delta\ln k$, and where $P(k)$ is the linear power spectrum defined as
\beq
P(k)=\average{|\delta_{\bo{k}}|^2}\ ,
\eeq
and $\delta_{\bo{k}}$ are the Fourier coefficients of the matter over-density field.

By numerically minimizing equation (\ref{eq:minimise}) over $z$ and
$s$ one finds a rescaling such that the linear variance of the
simulations are as similar as possible to each other across the range of
scales that correspond to the mass range of the haloes in the
original simulation. This is equivalent to minimizing the difference
in halo mass function because, in the simplest models, the mass function depends
only on $\sigma$ (\citealt{Press1974}; \citealt{Sheth1999}) as shown in the mass function in
equation (\ref{eq:STmf}). However, in more complicated models, such as those with
collapse thresholds that depend on environment (\eg \citealt{Mo1996}), 
this is no longer the case -- note also that strong environmentally dependent mass functions
are the case for most modified gravity theories (\eg \citealt{Lombriser2013}).

The result of this exercise has the issue that the desired
value of $z$ will almost certainly not be one of the values stored
as a simulation output; alternatively, each stored value of $z$ can be
assigned a corresponding $z'$, none of which will be exactly the desired
target value. In practice, this is not too important: simulation
outputs are used to build mock data on a light cone, which always
involves some degree of interpolation between outputs. The main thing
is that the grid of effective $z'$ values is known.
The problem can be eased if the
outputs from the original simulation are finely spaced in redshift. It can
also be an advantage to run this simulation with a high
value of $\sigma_8$ or alternatively into the
future (negative redshifts) in order to produce a large range in
fluctuation amplitudes, as this allows the algorithm to find scalings
between different cosmological parameters more easily 
(\eg \citealt{Harker2007}, AW10, \citealt{Ruiz2011}).

It may also be the case that, after remapping, the lowest mass
halo in the simulation is too massive to allow generation of
a realistic galaxy population. This is a problem with most simulations,
where the parent haloes of dwarf galaxies lie below the resolution
limit. In all cases, a reconstruction algorithm is 
thus required, in which the distribution of missing low-mass haloes
is inferred from the distribution of the known haloes
(\eg \citealt{delatorre2013}).

In Fig. \ref{fig:mf} we illustrate both the theoretical and measured
mass functions at various stages of the scaling process for two rather
different example cosmologies. This plot makes use of simulations
that are discussed in Section \ref{sec:simulations} and
summarised in Table \ref{tab:simulations}; briefly the two cosmologies
are a vanilla \lcdm model and \tcdm, a matter only model. Theoretical
agreement can be achieved almost perfectly (within 1\%) by rescaling,
but in the measured mass function there remains some disagreement at
around the 5\% level. A similar level of disagreement in the measured mass
function was found by AW10 (their Fig. 7); this plausibly reflects the fact that the mass
function is not perfectly universal (\citealt{Tinker2008};
\citealt{Lukic2007}; \citealt{Manera2010}). We do not use the mass
functions of \cite{Tinker2008} because they were calibrated on haloes
found with a spherical over-density algorithm (for examples see
\citealt{Knebe2011}) and on a small range of \lcdm parameter
space. They also have explicit redshift dependence which violates
universality, rather than depending only implicitly on redshift via
$\sigma$ (equation \ref{eq:sigma}). 
In principle one could choose $s$ and $z$ to minimise the
difference in mass function directly between two different cosmologies
and use a more complex $z$-dependent prescription for the mass function, but we have taken
the simpler approach for the present paper.

\subsection{Matching the displacement field}
\label{sec:displacement}

The second part of the AW10 algorithm involves a shift in the
individual particle positions in the rescaled simulation so as to
reproduce the large-scale clustering of the target
cosmology. This is achieved by taking the linear displacement field in the
scaled original cosmology and then using the Zel'dovich Approximation
(\citealt{Zeldovich1970}; hereafter ZA) to perturb the particle or halo
positions: the phase of each mode is preserved, but the
amplitude is altered to match the target power spectrum.

\begin{figure}
\begin{center}
\includegraphics[width=60mm,angle=270]{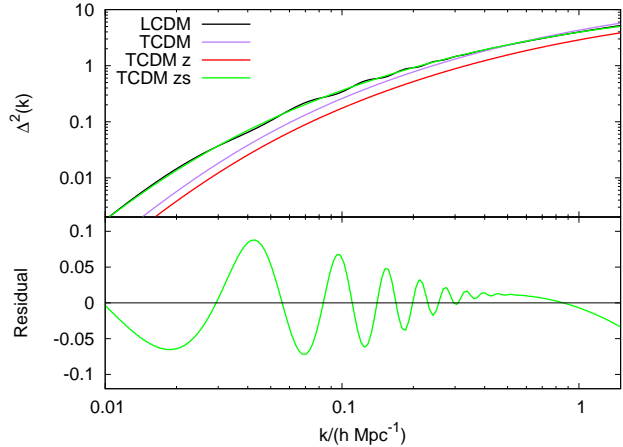}
\end{center}
\caption{The linear power spectrum of the original and target
  cosmologies described in Section \ref{sec:simulations} at each stage
  of the rescaling process. The black curve shows the target \lcdm
  simulation whereas the other lines show the various stages of the
  rescaling method: original \tcdm simulation (purple); scaling in
  redshift (red); and scaling in redshift and size (green). The
  residual difference in linear power after redshift and size scaling
  has taken place is shown in the lower panel, this is mainly obvious
  as the residual wiggle which arises because there is no BAO in the
  \tcdm model. This difference in linear power is
  corrected for by modifying particle positions, described in Section
  \ref{sec:displacement}.}
\label{fig:linear_power}
\end{figure}

Rescaling to match the halo mass function in effect forces the initial
simulation to take up the desired linear power spectrum in the region
with $\Delta_{\rm{lin}}^2\simeq1$. But in general the target spectrum will not be
matched on very different scales. This problem is illustrated clearly
in Fig. \ref{fig:linear_power}, where the target cosmology has BAO
features, whereas the original simulation adopted a zero-baryon
transfer function.
It is precisely these residual
differences in linear power that the next part of the algorithm aims
to correct by displacing particles using the ZA.

At each redshift in the target cosmology we define a nonlinear scale $R'_{\mrm{nl}}$ such that 
\beq
\sigma'(R'_{\mrm{nl}},z')=1\ ;
\label{eq:nl_scale}
\eeq 
all fluctuations on scales larger than this are considered to be
in the linear regime. AW10 then use this to define a nonlinear
wavenumber $k'_{\mrm{nl}}=R_{\mrm{nl}}^{'-1}$ that determines which
Fourier components of the density field and displacement field
are taken to be in the linear regime.

The displacement field $\bo{f}$ is defined so as to move particles
from their initial Lagrangian positions $\bo{q}$ to their Eulerian
positions $\bo{x}$: 
\beq 
\bo{x}=\bo{q}+\bo{f}\ .
\label{eq:q_to_x}
\eeq 
At linear order the displacement field is 
related to the matter over-density $\delta$ via 
\beq 
\delta=-\nabla\cdot\bo{f}\ , 
\eeq 
which in Fourier space is
\beq \bo{f}_{\bo{k}}=-i\frac{\delta_{\bo{k}}}{k^2}\bo{k}\ .
\label{eq:displacement}
\eeq
If the displacement field in the original simulation is known, 
then an additional displacement can be specified in Fourier space to 
reflect the differences in the linear matter power spectra 
between the two cosmologies:
\beq
\delta\bo{f}_{\bo{k'}}=\left[\sqrt{\frac{\Delta_{\rm{lin}}^{'2}(k',z')}{\Delta_{\rm{lin}}^2(sk',z)}}-1\right]\bo{f}_{\bo{k'}}\ ,
\label{eq:move}
\eeq
where $\bo{f}$ is measured in the original simulation after it has
been scaled. Equation (\ref{eq:displacement}) is only valid for the
linear components of both fields, so in practice the displacement field
must be smoothed with a window of width the nonlinear scale $R_{\mrm nl}$.

In AW10 the authors saved the initial displacement field of the
simulation and so equation (\ref{eq:move}) could be used directly to
make the required modification of the particle positions. 
But in the next Section we show how the displacement field can be
reconstructed directly from the distribution of haloes in the original
simulation.

\section{Recasting haloes}
\label{sec:recasting}

The AW10 algorithm produces a new particle distribution,  but many
practical applications would need to seed this density field with
galaxies, for which the first step is locating the dark matter
haloes. This takes time, and will also yield incorrect results since
the density field is not correct on the smallest scales (\ie the
internal halo properties should change as a result of the
altered cosmology). For both these reasons it
makes sense to work directly with the halo catalogue.
In this section, we therefore show both how the halo catalogue itself
can be used to recover the large-scale displacement field (if it is
not provided), and we review the changes to the internal halo
properties that should be applied after the simulation has been remapped.

\begin{figure*}
\subfloat{\includegraphics[width=60mm,angle=270,trim=0cm 3cm 0cm 3cm,clip,scale=1.25]{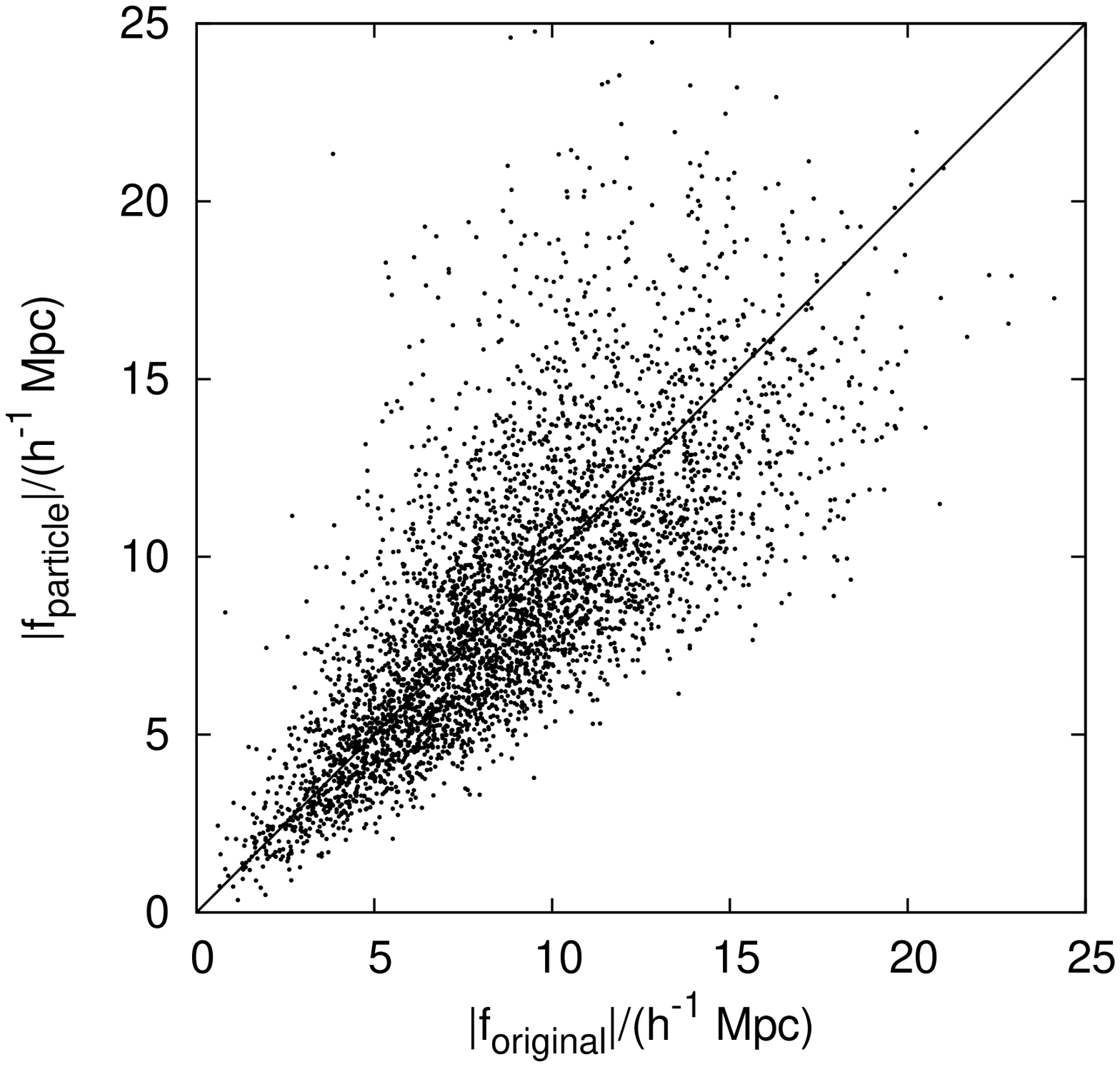}}\hspace{1cm}
\subfloat{\includegraphics[width=60mm,angle=270,trim=0cm 3cm 0cm 3cm,clip,scale=1.25]{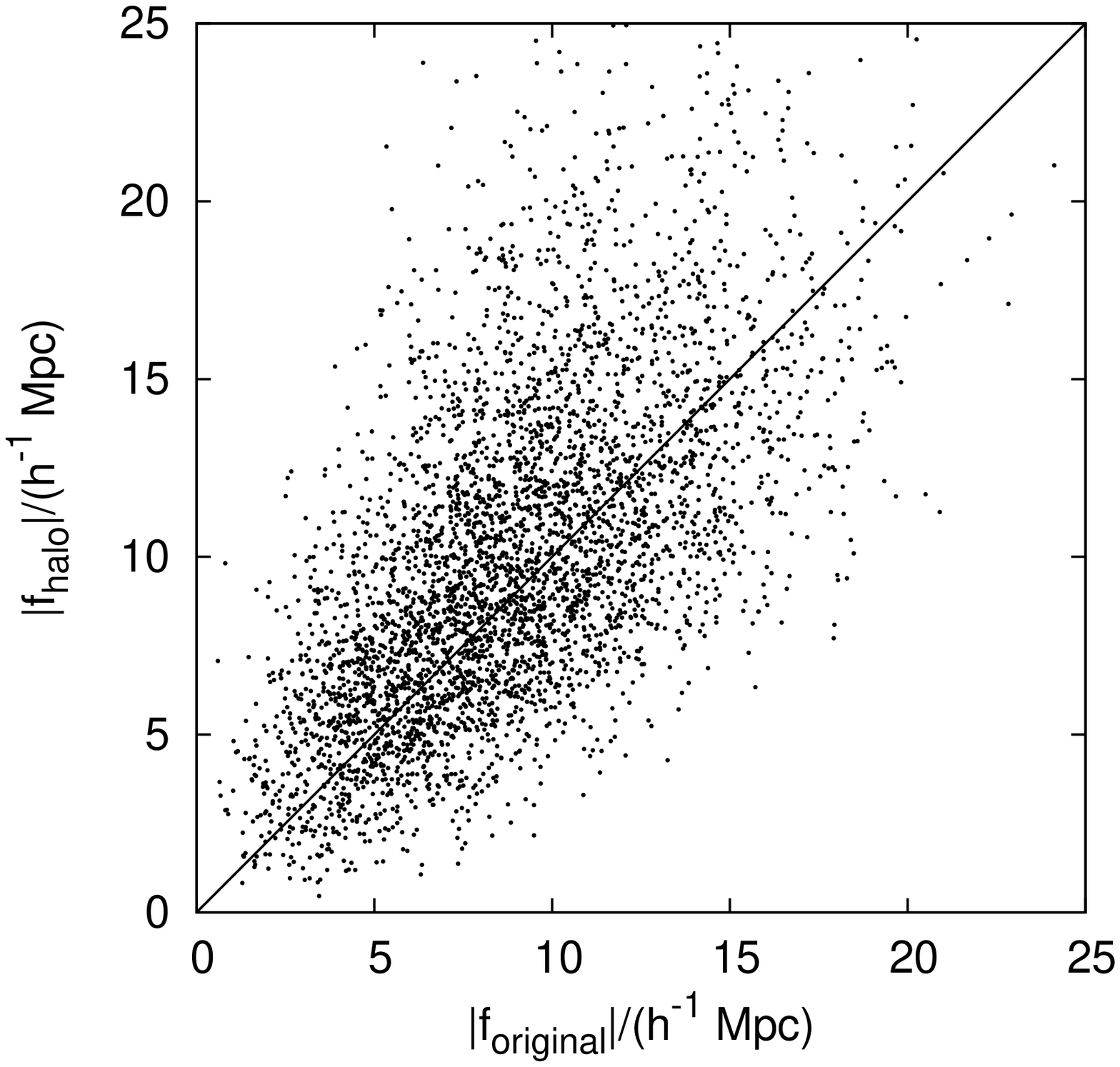}}\\
\subfloat{\includegraphics[width=60mm,angle=270,trim=0cm 3cm 0cm 3cm,clip,scale=1.25]{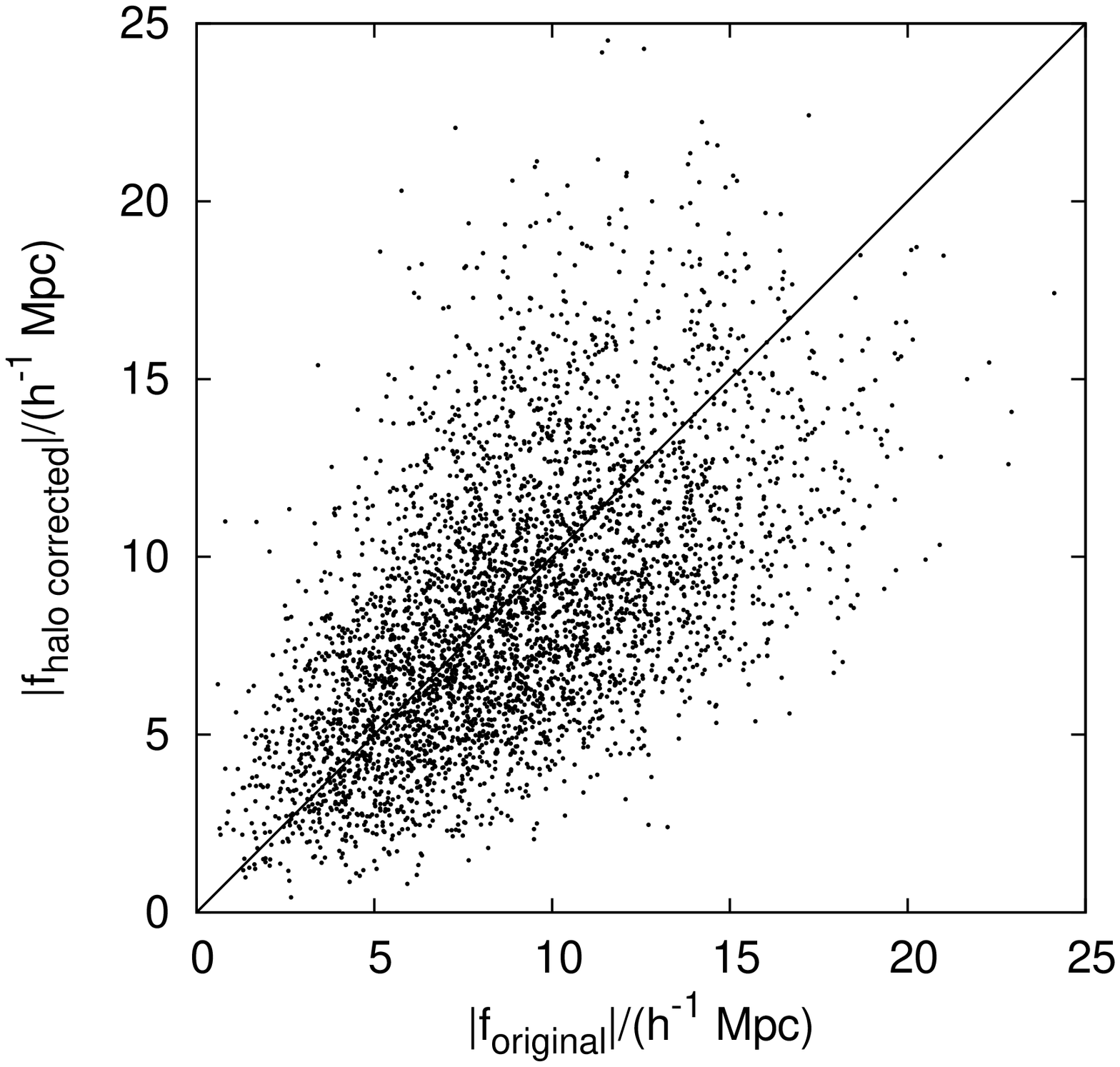}}\hspace{1cm}
\subfloat{\includegraphics[width=60mm,angle=270,trim=0cm 3cm 0cm 3cm,clip,scale=1.25]{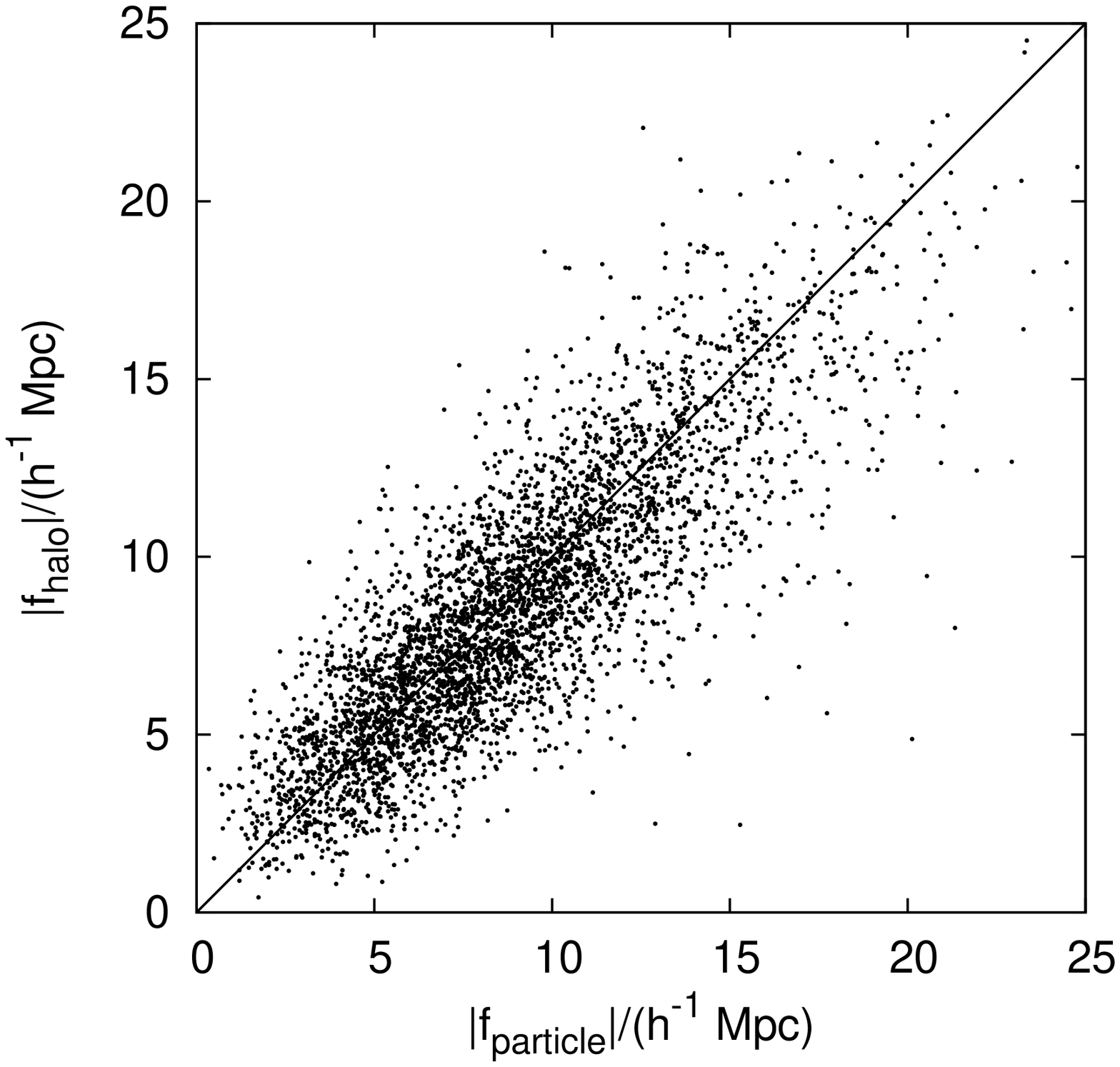}}\\
\caption{A comparison of the values of the linear displacement fields 
    reconstructed from our scaled (in $s$ and $z'$ by the first part of the method)
    \tcdm simulation (see Section \ref{sec:simulations}). The displacement
    field is calculated from the density field using equation
    (\ref{eq:displacement}) and debiased in the case of haloes using equation (\ref{eq:bias}).
    The points show values of the fields in cells for a randomly selected subset
    (1\%) of cells measured on a $75^3$ mesh and convolved with a
    Gaussian to filter out the nonlinear components.
    The top left panel shows the displacement reconstructed from the particles compared
    to the original displacement field used to run the simulation. The top right panel shows
    the same thing but for the displacement field recovered from the debiased halo field. 
    The lower left panel shows the halo displacement field corrected according to equation
    (\ref{eq:fix}) so as to have the correct theoretical variance. The bottom right panel
    shows a comparison between this corrected displacement field from haloes and the 
    field from the particles.}
\label{fig:scatter}
\end{figure*}

\subsection{Reconstruction of displacement fields}

Following \cite{Eisenstein2007}, the displacement field
can be obtained from the over-density field using equation
(\ref{eq:displacement}). This result can be used if we construct the matter
over-density field from the haloes, which are biased tracers of the
mass distribution. The over-density of haloes $\delta_{\mrm H}$ is related to
the matter over-density via the bias $b$: 
\beq 
\delta_{\mrm H}=b\delta\ ,
\label{eq:bias}
\eeq 
where the bias can, in principle, be a function of mass and other
halo properties.

Throughout this work we use the mass function of \cite{Sheth1999}:
\beq
f(\nu)=A\left[1+\frac{1}{(q\nu^2)^p}\right]e^{-q\nu^2/2}\ ,
\label{eq:STmf}
\eeq
where $A=0.216$, $q=0.707$ and $p=0.3$. 
The mass function is expressed
in terms of the variable $\nu=\delta_{\mrm c}/\sigma(M)$ where
$\delta_{\mrm c}\simeq 1.686$, which derives from spherical collapse models.  
$f(\nu)$ is defined such that
it gives the fraction of the mass in the Universe in haloes in a range
$\nu$ to $\nu +d\nu$. 
Although more up to date mass
functions exist in the literature (\citealt{Warren2006};
\citealt{Peacock2007}; \citealt{Tinker2008}) we choose to use that of
\cite{Sheth1999} because it was calibrated to simulations that cover
a greater range of cosmological parameter space than more modern ones.

Given the mass function, an analytic expression
for the linear halo bias can be derived via the peak-background split
formalism (\citealt{Sheth1999}): 
\beq
b(\nu)=1-\frac{1}{\delta_{\mrm c}}-\frac{\nu}{\delta_{\mrm c}}\left[\frac{d}{d\nu}\ln f(\nu)\right]\ .
\label{eq:STbias}
\eeq
In order to calculate the over-density field from our halo catalogue
we take a halo-number weighted `effective' bias for the haloes in the
catalogue based on the theoretical models given above in equations
(\ref{eq:STmf}) and (\ref{eq:STbias}): 
\beq
b_{\mrm{eff}}=\frac{\int_{\nu_{\mrm{min}}}^{\nu_{\mrm{max}}} \mathrm{d}\nu\,
  b(\nu)f(\nu)/m}{\int_{\nu_{\mrm{min}}}^{\nu_{\mrm{max}}} \mathrm{d}\nu\,
  f(\nu)/m}\ ,
\label{eq:eff_bias}
\eeq
where $\nu_{\mrm{min}}$ and $\nu_{\mrm{max}}$ are the value of $\nu$ for the least massive and most massive 
halo in the original simulation. 



Nonlinearities in the recovered matter over-density field 
are limited by convolving the
field with a Gaussian whose width is set equal to the nonlinear scale
$R_{\mrm{nl}}$, which can then be converted to a displacement field
using equation (\ref{eq:displacement}). Our method then proceeds
exactly as in AW10: haloes in the original simulation are
moved from their old positions $\bo{x}$ to new positions $\bo{x'}$
using the small displacements implied by 
equation (\ref{eq:move})
\beq
\bo{x'}=\bo{x}+\delta\bo{f}\ ,
\eeq
which follows from equation (\ref{eq:q_to_x}) given that initial positions $\bo{q}$
are preserved before and after this final stage of the algorithm.
In Fig. \ref{fig:scatter} we show the displacement fields as
predicted from the particle data and from halo catalogues in our
simulation (see Section \ref{sec:simulations}). The top left panel shows
a comparison between the displacement field reconstructed from the particle
distribution with that generated for the simulation initial conditions and we
can see that the reconstructed field shows no obvious bias compared to the original
fields although there is some scatter. 
The top right panel then shows the displacement field measured from
debiasing the halo density field which shows a small residual bias when compared to the original field.
This residual effect possibly reflects a failure of the
peak-background bias calculation in the quasi-linear regime.
In any case, though, the true expected variance in the smoothed
displacement can be calculated:
\begin{equation}
\sigma^2_f(R_{\mrm nl})=\frac{1}{3}\int_{k_\mrm{box}}^{\infty}\frac{e^{-k^2 R_{\mrm nl}^2}\Delta^2_{\mrm{lin}}(k)}{k^2}\;\mrm{d}\ln k\ ,
\label{eq:sigv}
\end{equation}
where $k_\mrm{box}=2\pi/L$ is the fundamental mode of the simulation.
We can therefore scale our
displacement fields such that they have the desired
variance:
\begin{equation}
\bo{f} \rightarrow \bo{f} \, \frac{\sigma_f(R_{\mrm{nl}})}{\sqrt{\mrm{Var}(|\bo{f}|)}}\ ,
\label{eq:fix}
\end{equation}
where $\mrm{Var}(|\bo{f}|)$ is the measured variance in $|\bo{f}|$.
The result of this scaling can be seen in the bottom left panel
of Fig. \ref{fig:scatter} where there is now better agreement
with the original displacement field. The
bottom right panel shows a comparison between the reconstructed displacement
field from particles and from haloes where there is no obvious disagreement.
This shows that our method is able to make a reasonable reconstruction of the
full simulation displacement field using only the halo catalogue.

One should note that for a population of lower mass haloes the value of $b_{\mrm{eff}}$ 
could be less than 1 and an implementation of equation (\ref{eq:bias}) could then result in cells with negative
densities ($\delta<-1$). However, we have checked our reconstruction method for a small volume simulation
with a population of low mass haloes with $b_{\mrm{eff}}\approx 0.83$ and found that it still works as well
in reconstructing the displacement field, even though it goes through the unphysical negative density step.

\begin{figure*}
\includegraphics[width=110mm,angle=270,trim=1.8cm 1.8cm 1.8cm 1.8cm]{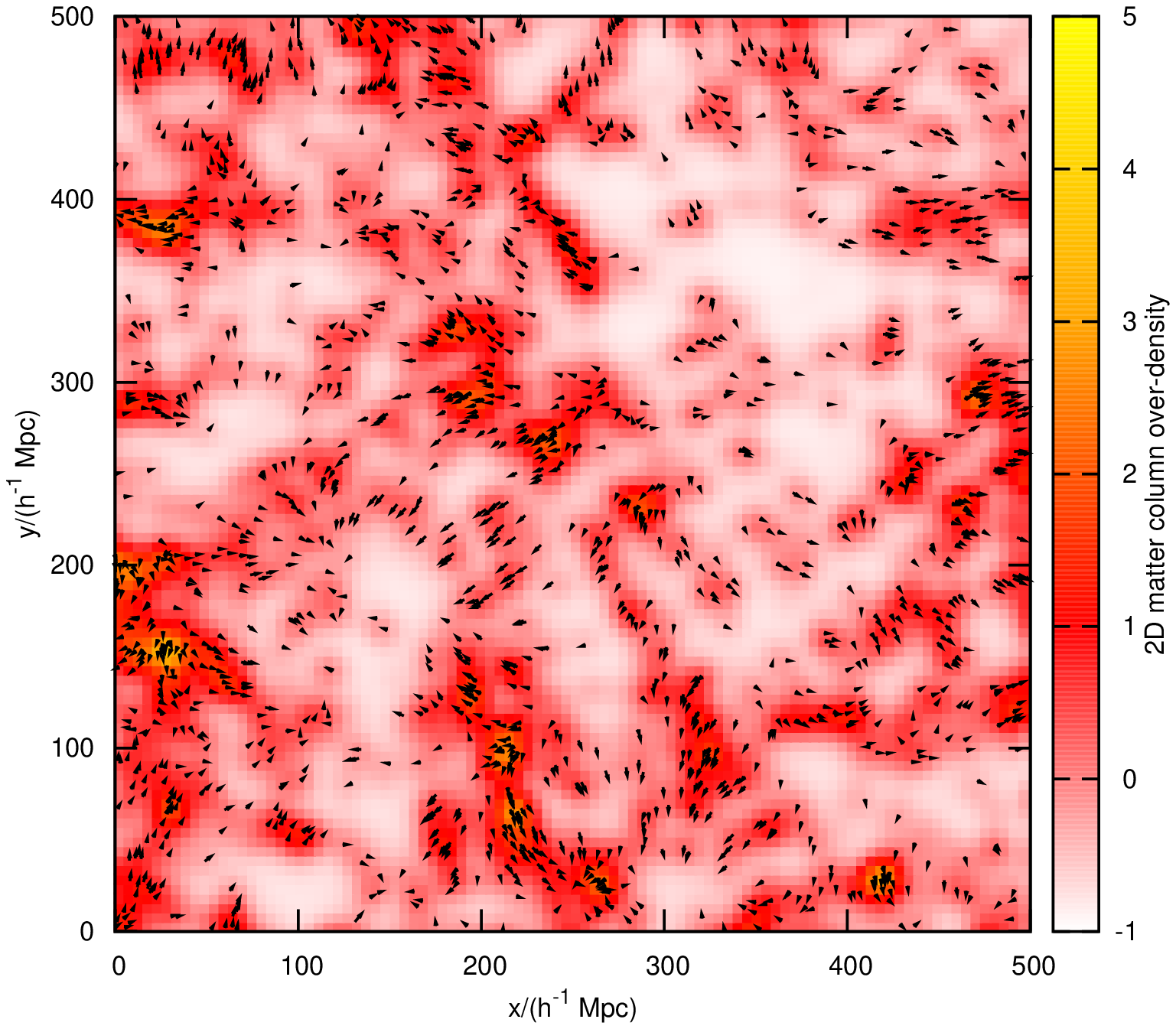}
\includegraphics[width=110mm,angle=270,trim=1.8cm 1.8cm 1.8cm 1.8cm]{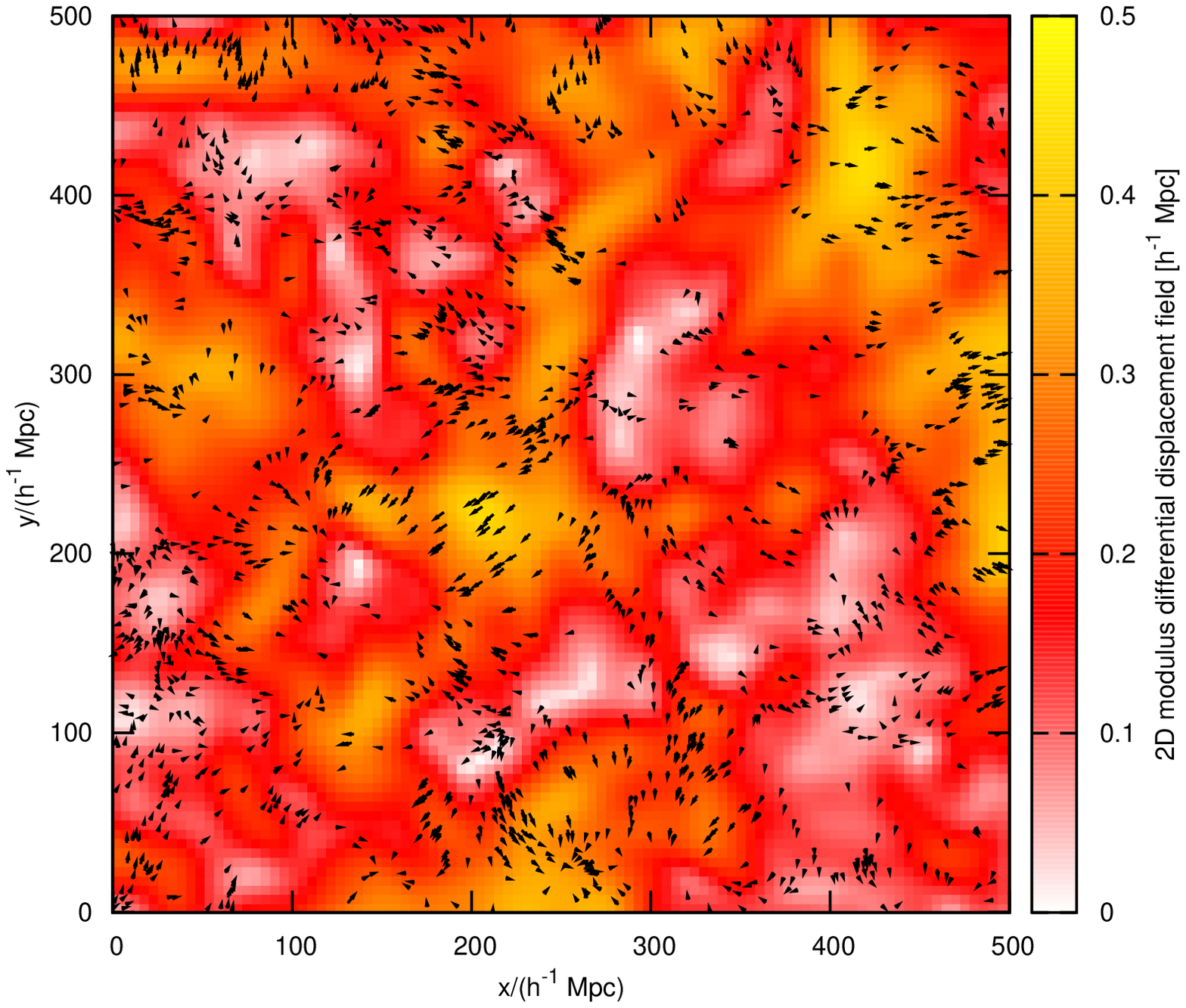}
\caption{A visual summary of our method. The top panel shows the
  projected linearized over-density field in a slice of thickness
  $50\Mpc$ inferred from the distribution of haloes in our size and redshift scaled \tcdm
  simulation (described in Section \ref{sec:simulations}) and the
  bottom panel shows the magnitude of the linearized differential
  displacement field inferred from the over-density. In each plot the
  arrows then show the flow of haloes due to the differential
  displacement field in order to match the clustering in the target \lcdm
  cosmology. The displacements are typically small in the method and
  the arrows in these plots have been enlarged by a factor of 10 to
  illustrate the halo flow more clearly.}
\label{fig:big}
\end{figure*}

\subsection{Mass-dependent halo displacements}
\label{sec:mass_bias}

When dealing with the displacement field of haloes,
some care is needed in ensuring that these objects
display the correct degree of bias as a function of mass. 
Writing the matter density fluctuation in terms of the
displacement field, the linear halo bias relation is
\beq
\delta_{\mrm{H}}=-b(M)\nabla\cdot\bo{f}\ ,
\label{eq:biasmove}
\eeq
which says that in effect haloes of different masses are displaced 
by different amounts. This seems to violate the 
equivalence principle, and of course all particles in a simulation
should share the same displacement field. But this displacement field
then affects halo formation in a nonlinear way, which is not allowed
for if we subsequently change the displacement field `by hand'.
In order to obtain the correct statistics
of large-scale clustering, the above mass dependence 
of the effective additional displacement must therefore be respected.
To see how the argument works in an extreme case, imagine applying the
AW10 method to a simulation with zero-large-scale power. Adding in the
large-scale displacement field will then by construction yield a
set of haloes that have $b=1$, independent of mass. In order to
avoid this unrealistic situation, we have to apply a mass-dependent
displacement, as in  equation (\ref{eq:biasmove}). 

This argument reveals a subtle limitation of the original AW10 prescription.
We can assume that applying a halo finder to a particle distribution
that has been subject to the AW10 method will find very much the
same haloes as if these were identified prior to the additional
displacement, because these displacements are coherent over large scales.
These haloes will thus fail to have the correct
dependence of clustering on mass. In this respect, our approach
is not simply faster than AW10, but working directly with haloes
allows a treatment of mass-dependent biasing that is more consistent
than can be achieved by scaling the particle distribution alone.

In practice one could bin haloes of differing masses and
compute the displacement field for each mass bin individually, thus
avoiding the issue of debiasing the over-density and then rebiasing
the displacement field. However we choose to use the full halo
catalogue to produce the least noisy displacement field possible 
and then to move haloes of different masses by different
amounts according to equation (\ref{eq:biasmove}). 

A visual summary of our method as applied to halo catalogues is given
in Fig. \ref{fig:big}, in which we show the density and displacement
fields as calculated from the halo distribution together with the flow
of haloes that these fields imply for our different cosmologies.

\subsection{Reconstitution of haloes}
\label{sec:reconstitution}

The AW10 method reproduces the mass function and linear clustering of
the target cosmology, albeit with the small error in mass-dependent
halo biasing described above. But in addition, the AW10 approach
does not address the deeply nonlinear
clustering that arises due to correlations within individual
haloes. In Halo Occupation Distribution (HOD) models, galaxies are
taken to be stochastic tracers of the mass field around haloes; in
order to use rescaling for generation of mock galaxy catalogues, it is
therefore necessary to produce the mass field around haloes in a way
that reflects the new cosmology. This is also of interest in its own
right for applications such as ray-tracing simulations
(\eg \citealt{Kiessling2011}) for gravitational lensing.

We address this issue by a method of `reconstitution' where
the mass distribution around the final set of haloes 
is calculated by considering how their
internal structure should depend on cosmology.
We define haloes as spherical objects that have an average
over-density with respect to the matter in the background Universe of
$\Delta_{\mathrm{v}}\simeq 200$. The use of a fixed density
contrast at virialization is motivated by consistency with halo finding
methods such as Friends-of-Friends (FoF).
The exact value of $\Delta_{\mathrm{v}}$ (motivated by the spherical
collapse model) is not critical.
The virial radius $r_{\mrm v}$ for a halo of mass $M$ is then 
\beq
r_{\mrm v}=\left(\frac{3M}{4\pi\Delta_{\mrm v}\bar{\rho}}\right)^{1/3}\ .
\label{eq:virial_radius}
\eeq

Haloes differ between cosmologies in having different density profiles;
this can be traced to the haloes having different collapse
redshifts, via the differing growth rate of
perturbations (\citealt{Navarro1997}; \citealt{Bullock2001};
\citealt{Eke2001}). Although we have defined haloes to have a
fixed virial radius for a given mass, the concentration
of haloes (ratio of virial radius to internal characteristic radius) 
does vary as a function of cosmology. The density profiles
of haloes have been claimed to be universal (\citealt{Navarro1997}) with
a functional form of
\beq
\rho(r)=\frac{\rho_{\mrm N}}{(r/r_{\mrm s})(1+r/r_{\mrm s})^2}\ ,
\label{eq:nfw}
\eeq
where $r$ is the distance from the halo centre, $r_{\mrm s}$ is a scale
radius and $\rho_{\mrm N}$ is a normalization to obtain the correct halo
mass. Although subsequent work (\citealt{Merritt2005};
\citealt{Merritt2006}) showed this form to be imperfect at small
$r$, it will suffice for our present purpose: mock galaxies are
either central at $r=0$ exactly, or satellites that tend to be
found around $r_{\mrm s}$, where the NFW approximation is good.
The density
profile is truncated at the virial radius $r_{\mrm v}$, which is
determined by the halo mass (equation \ref{eq:virial_radius}); the
density profile is therefore fully specified via a value
for $r_{\mrm s}$ or alternatively for the halo concentration $c=r_{\mrm v}/r_{\mrm s}$. 

We choose to use the cosmology dependent concentration relations of
\cite{Bullock2001} for convenience; these
are defined as a function of redshift $z$ to be 
\beq
c(z,z_{\mrm c})=4\left(\frac{1+z_{\mrm c}}{1+z}\right)\ , 
\eeq 
where $z_{\mrm c}$ is a collapse
redshift which depends on halo mass, calculated via 
\beq
\frac{g(z_{\mrm c})}{g(z)}\,\sigma(0.01M,z)=\delta_{\mrm c}\ , 
\eeq 
where $g(z)$ is
the linear growth factor normalized to $1$ at $z=0$. This expression gives the
redshift at which the halo has gathered $1\%$ of its current
mass. Since both $\sigma$ and $g$ depend on cosmology, the halo
concentrations depend on cosmology, with more massive haloes collapsing later
and being consequently less concentrated. The concentration relations of
\cite{Bullock2001} were calibrated using haloes whose virial radius
was defined to vary as a specific function of cosmology according to the spherical model
approximation of \cite{Bryan1998}.
\beq
\Delta_{\mrm v}^{\mrm B}(z)=\frac{178-82[1-\Omega_{\mrm m}(z)]-39[1-\Omega_{\mrm m}(z)]^2}{\Omega_{\mrm m}(z)}\ .
\eeq
As argued above, this may not be the correct choice in detail when using 
FoF haloes with a cosmology-independent linking length, and the appropriate
value is probably best measured empirically in a given simulation (see Section \ref{sec:velocities}).
In this case one
should modify the \cite{Bullock2001} value for the concentration at given mass:
\beq
\left[\ln(1+c)-\frac{c}{1+c}\right]=\frac{\Delta_{\mrm v}}{\Delta_{\mrm v}^{\mrm B}}
\frac{c^3}{c_{\mrm B}^3}\left[\ln(1+c_{\mrm B})-\frac{c_{\mrm B}}{1+c_{\mrm B}}\right]\ .
\eeq 
Unless otherwise stated, we adopt $\Delta_{\mrm v}=200$ as a default value. This choice
is not critical, since we are often interested in differential effects
between cosmologies, and the main cosmology dependence of halo properties
comes through the influence on the concentration of altered formation redshifts.

\begin{figure}
\begin{center}
\includegraphics[width=70mm,angle=270,trim=0cm 2.5cm 0cm 0cm]{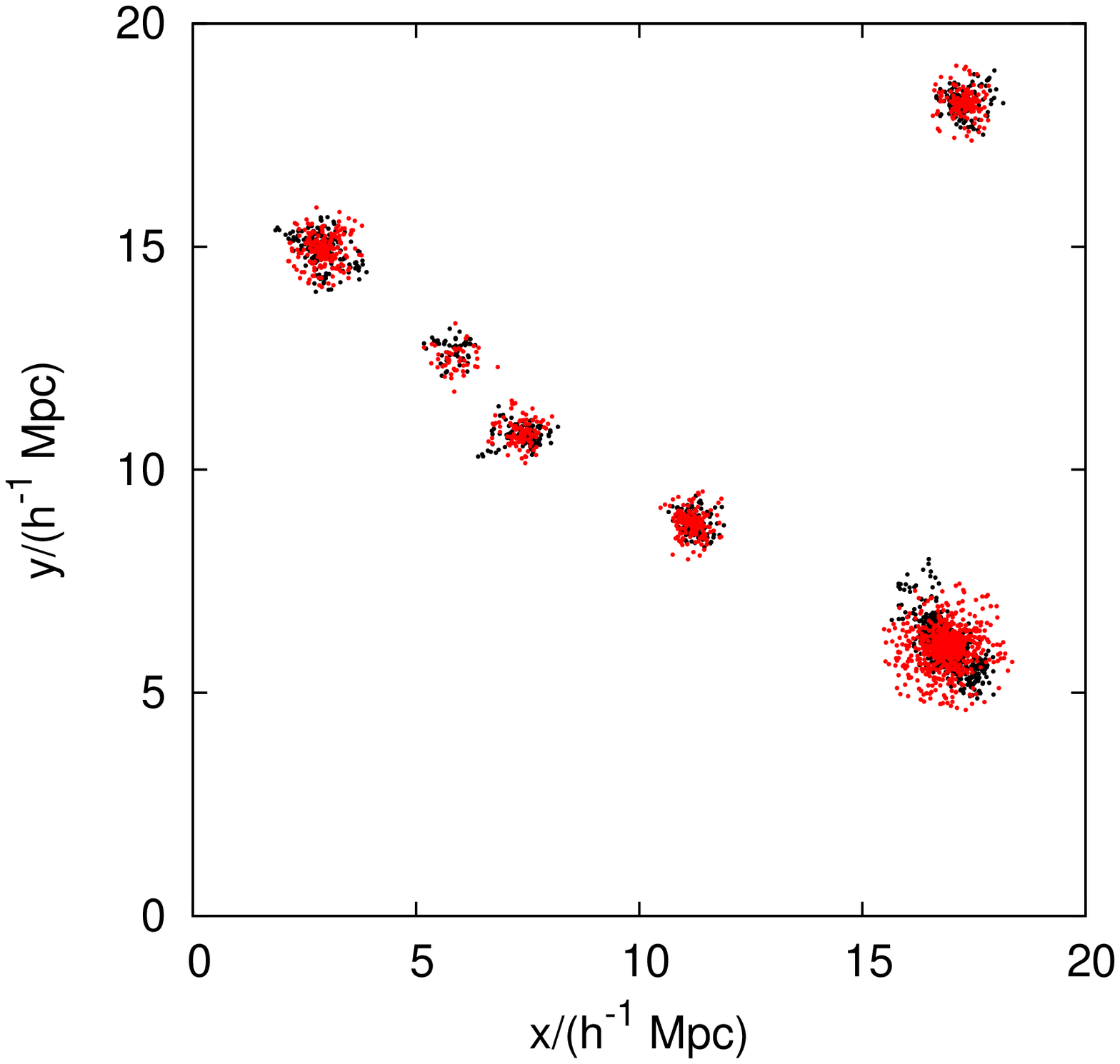}
\includegraphics[width=70mm,angle=270,trim=0cm 2.5cm 0cm 0cm]{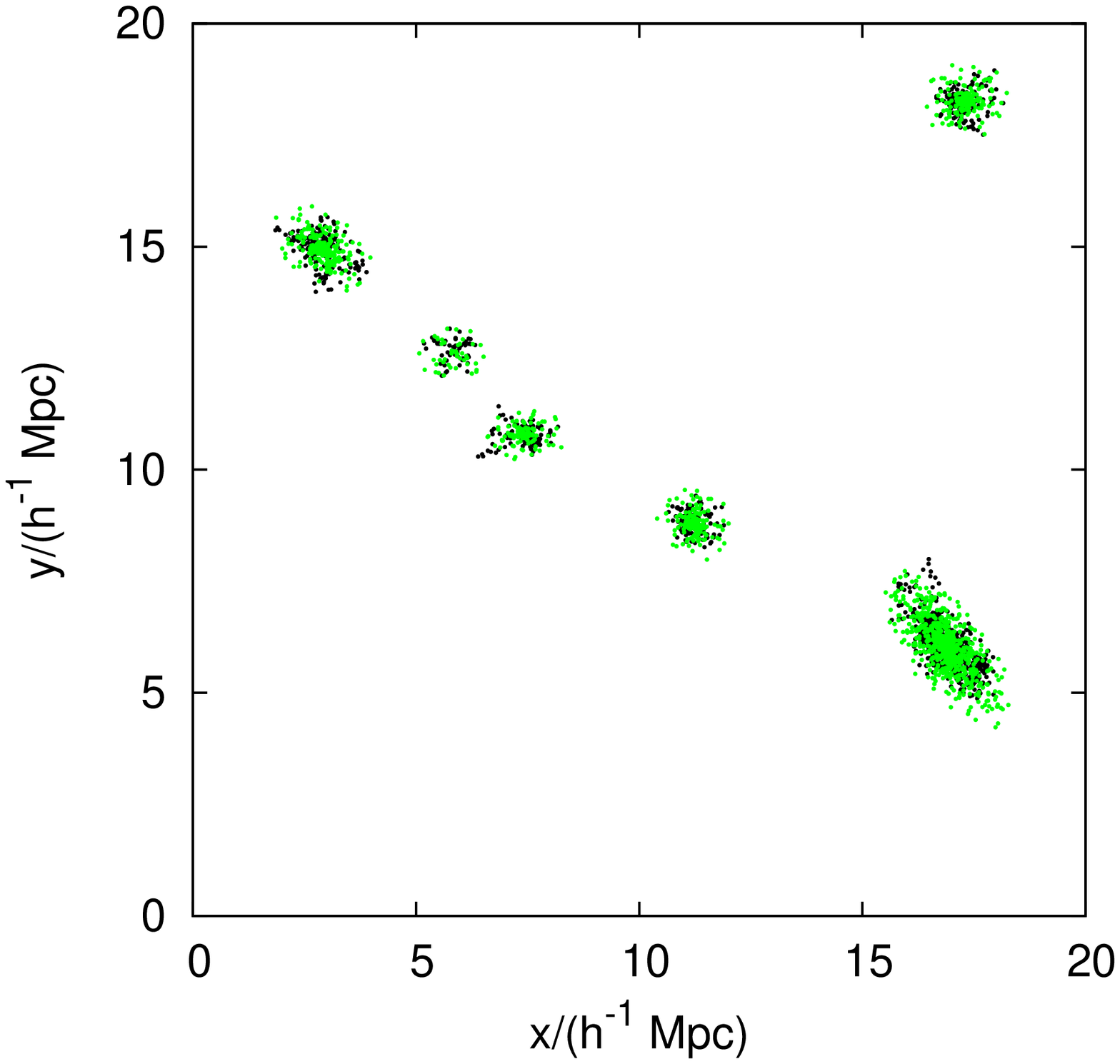}
\end{center}
\caption{The particles in a small cube of the \tcdm simulation of side
  length $20\Mpc$ that were grouped into haloes by our FoF algorithm
  are shown in black. In the top panel overplotted in red are our
  reconstituted spherical NFW haloes; in the bottom panel our
  reconstituted aspherical NFW haloes are shown in green, and these
  clearly match the haloes in the simulation much better.}
\label{fig:reconstitution}
\end{figure}

One can use the information in this section to reconstitute the
particles contained in each halo by calculating the virial radius and
concentration parameter for each halo in the catalogue and then
filling up the density profile around the halo by a random sampling of
tracer particles which correspond to those in the original
simulation. A pictorial representation of this is shown in
Fig. \ref{fig:reconstitution} where haloes measured in a simulation
with a FoF algorithm (see Section
\ref{sec:simulations}) are shown together with those reconstituted
using the halo catalogue generated from this distribution.

The top panel of Fig. \ref{fig:reconstitution} shows that `real'
haloes are often far from spherical, so it is better to reconstitute
them as triaxial objects. This is done using the moment of inertia
tensor 
\beq
I_{ij}=\sum_{k=1}^{N}(x_{k,i}-\bar{x}_{i})(x_{k,j}-\bar{x}_{j})\ ,
\eeq 
where $k\in\{1,...,N\}$ and there are $N$ particles in each halo
and $i,j\in\{1,2,3\}$ and label coordinates. We work with haloes
of 100 particles or more which we consider to be adequate for estimating
this tensor. Diagonalizing this tensor
provides the axis ratios of the halo (via the eigenvalues) and the
orientation of the halo (via the eigenvectors). The eigenvalues and
eigenvectors are stored when we generate the halo catalogue from the
particle distribution. Asphericity is then restored to the haloes by
distorting them once they have been generated by the spherical halo
reconstitution process described above. If the square roots of the
eigenvalues are $a$, $b$ and $c$ then each coordinate of the
reconstituted halo in the centre of mass frame is modified according
to
\begin{eqnarray}
x_i &\rightarrow& 3ax_i/(a+b+c)\ , \nonumber \\
y_i &\rightarrow& 3by_i/(a+b+c)\ , \nonumber \\
z_i &\rightarrow& 3cz_i/(a+b+c)\ .
\end{eqnarray}
We also considered the prescription $x \rightarrow ax/(abc)^{1/3}$ \etc
but found this not to work as well in recovering the shapes of aspherical haloes.
The CM position vector of each halo particle is then rotated by the
inverse matrix of eigenvectors in order to orient the halo correctly.

\begin{figure}
\begin{center}
\includegraphics[width=60mm,angle=270]{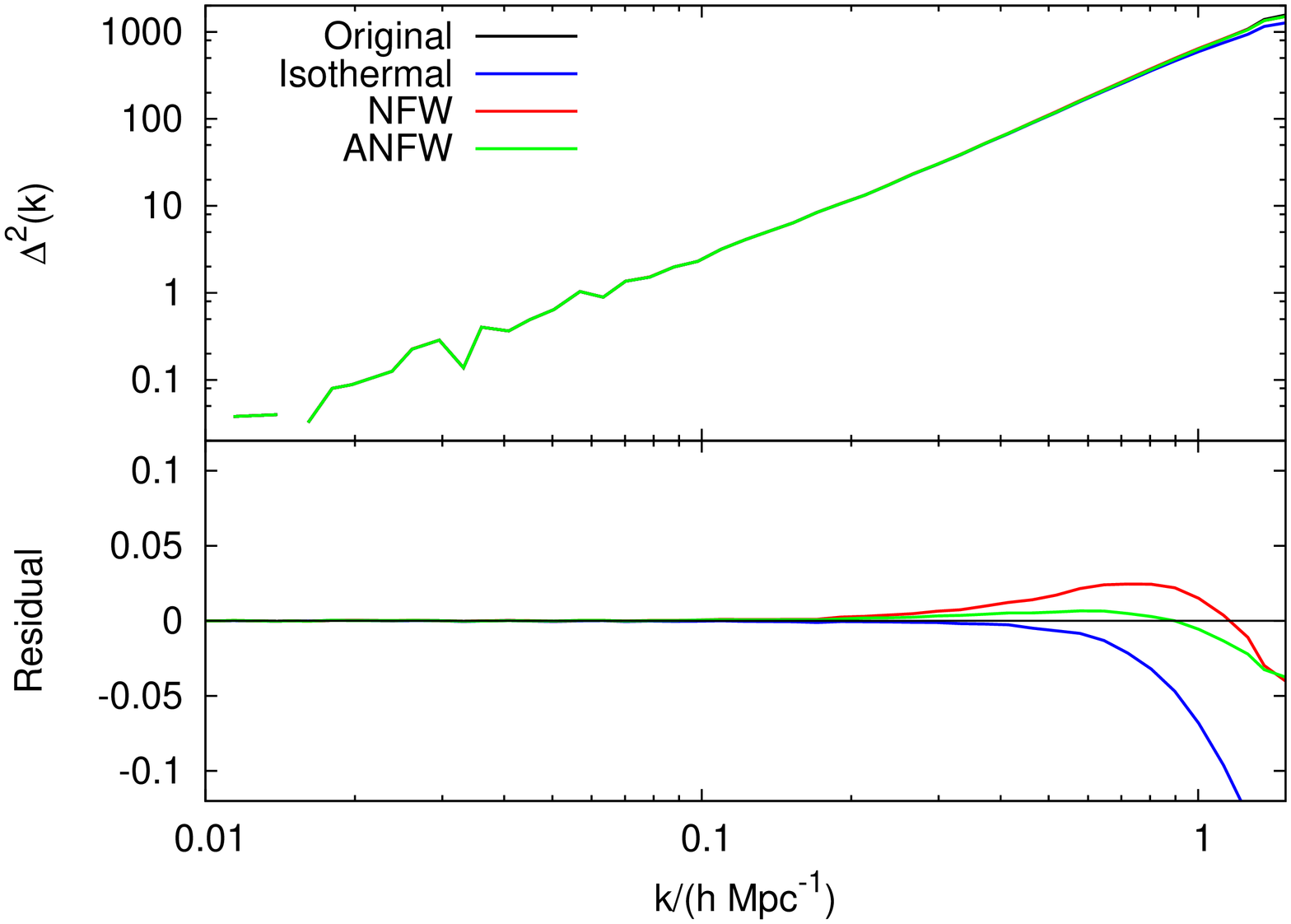}
\includegraphics[width=60mm,angle=270]{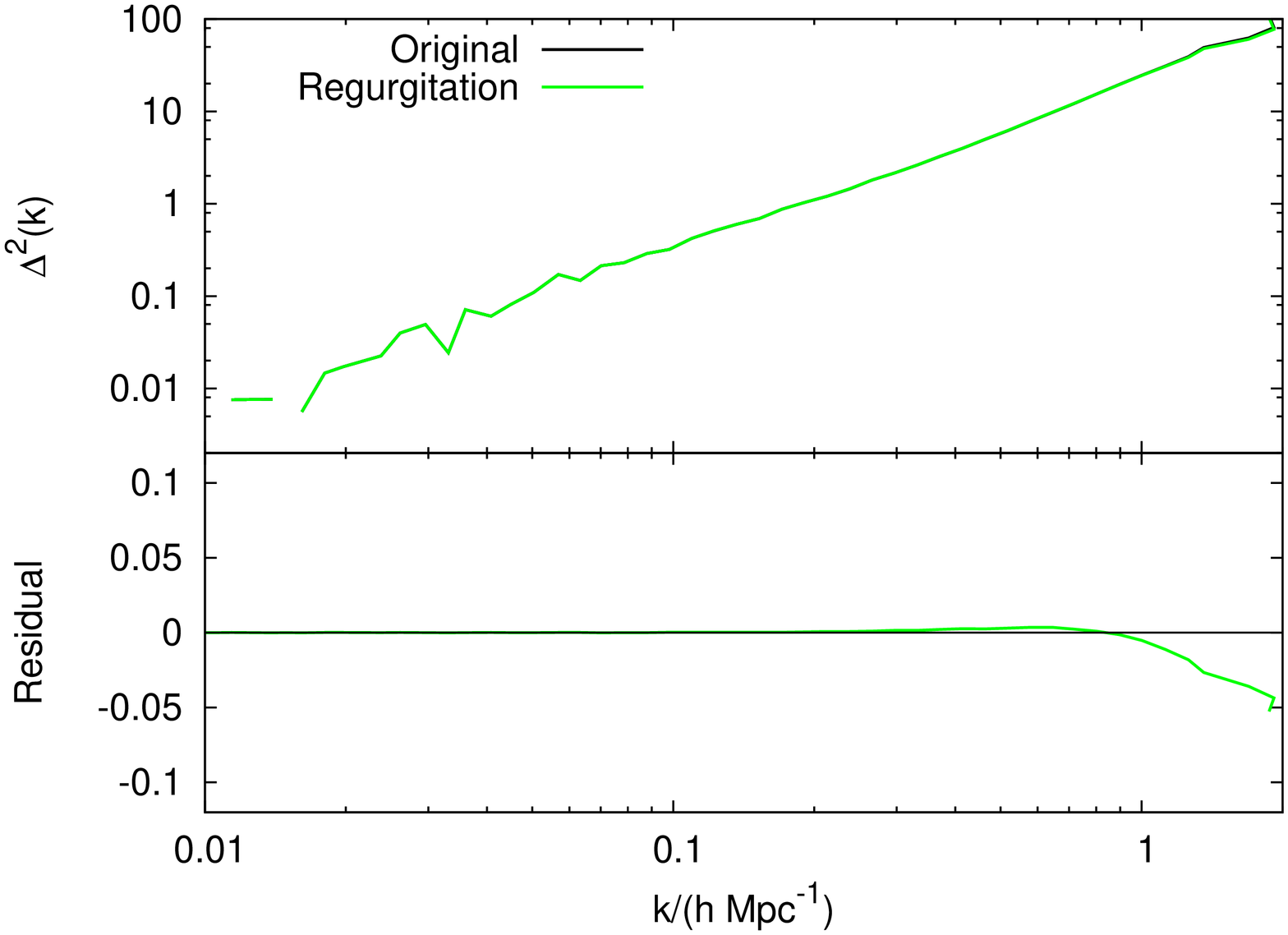}
\end{center}
\caption{The top panel shows the power spectrum of particles in
  haloes. We show spectra of the particles in the original haloes in
  our \lcdm simulation (black) together with the power of spherical
  (red) and aspherical (green) NFW haloes generated from the halo
  catalogue via the halo reconstitution process described in the
  text. There is  a clear improvement in the match of the power spectra gained from
  using aspherical rather than spherical haloes. 
  Also shown for comparison is the power
  spectrum for a more unrealistic halo: the singular isothermal sphere (blue);
  this fails to match the target even at
  fairly large scales. 
  In the lower panel we show the full matter
  power spectrum after we have regurgitated our reconstituted haloes
  back into the parent particle distribution. One can see the match is
  essentially perfect to around $k=1\iMpc$. The 5\% drop in power at
  smaller scales in both panels may reflect either imperfect
  concentration relations or lack of halo substructure in our
  reconstitution.}
\label{fig:perfect_recon_halo_power}
\end{figure}

In the top panel of Fig. \ref{fig:perfect_recon_halo_power} we show
the power spectrum of the particles in haloes after they have been
reconstituted from a halo catalogue, and we compare this to the power
spectrum of the particles in haloes in the original simulation that was
used to create the catalogue. Clearly the
clustering will agree on large scales, but it is satisfying to see that
the power spectrum of the particles in haloes can be reproduced by
generating our own NFW haloes even out to relatively small scales
($k\simeq1\iMpc$). One can see that there is also a significant
improvement in the matching in the clustering gained by reconstructing
aspherical haloes rather than purely spherical ones.

The final idea we consider is a method of `regurgitation', in order to
recreate the full mass distribution in the best possible way.
Once the original AW10 algorithm has been applied to a full
particle distribution, haloes are then selected and removed from the
particle distribution and then reconstituted in the same way as
described above. These reconstituted haloes, with the correct internal
structure for the new cosmology, are then reinserted into the
rescaled mass distribution in order to produce a corrected full
particle distribution for the new cosmology. In doing this we avoid
the problem of discontinuities between the reconstituted halo and the
surrounding material by using a constant $\Delta_{\mrm v}$ for our haloes so
that they have identical virial radii independent of cosmology. 
One should note that a limitation of this approach is that we move
all particles in the simulation according to the same displacement field and so
haloes are not subject to the biased displacements discussed in Section 
\ref{sec:mass_bias}. This is a general limitation of the AW10 method when one deals with the
particle distribution rather than the halo distribution.

The lower panel of Fig. \ref{fig:perfect_recon_halo_power} shows the
full matter power spectrum measured in a perfect test case where no
rescaling has taken place. Haloes have been identified with a FoF
algorithm and removed from the particle distribution. This halo
catalogue is then used to reconstitute haloes and these are then
regurgitated back into the surrounding particle distribution of the
simulation. The power spectrum is able to be recreated perfectly up to
$k=1\iMpc$ where deviations arise, possibly due to lack of
substructure or imperfect concentration relations in the reconstituted
haloes. It should be possible to improve this situation by using the
exact 3D particle distribution of the haloes, and scaling radii according
to the different concentrations in the two cosmologies. We have not
pursued this in detail here, since our focus is on seeing how much
can be achieved using halo catalogues alone.

\subsection{Scaling velocities}
\label{sec:velocities}

All the discussion so far has been in configuration space. But
galaxy surveys inhabit redshift space, in which the clustering
signature is modified by peculiar velocities. This distortion
is well-known to be an invaluable source of additional cosmological
information, giving direct access to the growth rate of
density perturbations (\citealt{Kaiser1987}; \citealt{Reid2012}).
We therefore now give a discussion on
how to scale particle and halo velocities.

The main element of scaling of velocities in cosmological
simulations is explained in Section 15.7 of \cite{b:peacock}. 
Since a computational volume has no knowledge of the physical
size that it is intended to represent, the natural
measure of velocity is ${\bo U}\equiv {\bo v}/HaL$, \ie peculiar
velocity in units of expansion across the box (whose proper size is $aL$). 
According to the Zel'dovich approximation, $\bo U$ is equal to the
displacement field in units of the box size, times the
logarithmic growth rate $f_{\mrm{g}}\equiv \mrm{d}\ln\delta/\mrm{d}\ln a\simeq \Omega^{0.55}_\mrm{m}(a)$
(where the latter approximation applies for a flat $\Lambda$-dominated model).
In other words, for two simulations that have identical fluctuation spectra
in box units (which is exactly true by definition in reinterpreting an original 
simulation output), we would expect the value of $\bo U$ to be
unaffected by a change in cosmology, apart from the alteration in $f_{\mrm{g}}$.
The recipe for
rescaling large-scale peculiar velocities is thus
\beq
\bo{v}'=s\frac{H'f'_{\mrm{g}}a'}{Hf_{\mrm{g}}a}\bo{v}\ .
\eeq

This argument does not apply on small scales, where velocities
are due predominantly to bound motions in haloes. But the
error is not large: according to the `cosmic virial theorem'
of Section 75 of \cite{b:peebles}, the pairwise peculiar
velocity dispersion for a given level of mass clustering
scales as $\Omega_{\mrm{m}}^{0.5}$. Therefore, simply rescaling
the velocities according to linear theory would give a
result in error on small scales by only about 7\%, even
when rescaling from $\Omega_{\mrm{m}}=1$ to $\Omega_{\mrm{m}}=0.25$.

However, the above scaling
does not account for the large-scale modifications to displacement
fields as discussed in Section \ref{sec:displacement}. 
in the Zel'dovich approximation, peculiar
velocities are assigned to particles by
\beq
\bo{v}=aHf_{\mrm g}\bo{f}\ ,
\eeq
we can therefore impose additional differential changes on the peculiar velocities of particles via
\beq
\delta\bo{v}_\bo{k'}=a'H'f_{\mrm g}'\left[\sqrt{\frac{\Delta_{\rm{lin}}^{'2}(k',z')}{\Delta_{\rm{lin}}^2(sk',z)}}-1\right]\bo{f}_\bo{k'}\ .
\eeq
In this and the earlier discussion, it should be kept in mind that the
velocities are in proper units, but ${\bf f}$ is a comoving displacement
field; this accounts for the extra factor of $a$.
Note that this additional velocity is applied independent of halo mass,
unlike the mass-dependent displacement discussed earlier. The latter
step was needed to preserve the mass-dependent biasing, but velocities
of haloes have no such mass dependence. Therefore, in effect, it is 
necessary to break the Zel'dovich approximation in order to ensure
correct large-scale statistics.

In addition, the velocity of particles in haloes will be different in
the new cosmology. The circular velocity at a radius $r$ from the
centre of a halo of mass $M$ can be calculated for an NFW profile:
\beq
v_{\mrm{circ}}^2(r)=\frac{GM}{r}\left[\frac{\ln(1+r/r_{\mrm{s}})-\frac{r/r_{\mrm{s}}}{1+r/r_{\mrm{s}}}}
{\ln(1+c)-\frac{c}{1+c}}\right]\ ,
\eeq
where $G$ is the gravitational constant. If the halo particles themselves are
restructured, then velocities can be assigned to particles in haloes in
a differential way:
\beq
\bo{v'}=\bo{v}\frac{v'_{\mrm{circ}}(r')}{v_{\mrm{circ}}(r)}\ .
\eeq
More normally, we might lack any internal halo velocity data, in which case the
velocities would need to be generated by hand. The simplest approximation would
be to assume isothermal and isotropic orbits; this is not consistent, and
more detailed modelling could be carried out based on the Jeans Equation, together with assumptions
about orbital anisotropy. But for the present, we shall go no further than noting
that virial equilibrium and isotropy yields an rms line-of-sight velocity dispersion for an NFW
halo of
\beq
\eqalign{
\sigma_v^2 &= {GM\over 3r_{\mrm v}}\; {c[1-1/(1+c)^2-2\ln(1+c)/(1+c)]
\over 2[\ln(1+c) - c/(1+c)]^2} \cr
&\simeq \left[\frac{2}{3}+\frac{1}{3}\left(\frac{c}{4.62}\right)^{0.75}\right]\frac{GM}{3r_{\mrm v}}\ .}
\label{eq:nfw_dispersion}
\eeq
This can be compared with $\sigma_v^2 = GM/3r_{\mrm v}$ for the truncated singular isothermal sphere.
We found that equation (\ref{eq:nfw_dispersion}) with $\Delta_{\mrm v}=200$ under-predicts
halo velocity dispersions in $\Lambda$CDM simulations by a factor of
around 1.07, implying that $\Delta_{\mrm v}\simeq 300$ would be a better practical choice for this application.
Better still, if a halo velocity dispersion is included in a catalogue, then  a scaled version of this
value can be used directly to reconstitute halo particles with the correct dispersion.

\subsection{Method summary}

Here we provide a brief summary of a practical implementation of our method for use on a halo catalogue:

\begin{enumerate}
  \item Calculate $z$ and $s$ by minimizing equation (\ref{eq:minimise}) over the mass range of haloes in the original halo catalogue.
  \item Calculate the effective bias for the haloes using equation (\ref{eq:eff_bias}).
  \item Calculate the matter over-density field implied by the halo catalogue, taking care to debias the halo field appropriately.
  \item Linearize the matter over-density field using a Gaussian with width of the nonlinear scale, defined in equation (\ref{eq:nl_scale}).
  \item Compute the displacement field from the over-density field using equation (\ref{eq:displacement}) and then correct this so that it has the correct theoretical variance using equation (\ref{eq:fix}).
  \item Taking the original catalogue at redshift $z$, relabel positions of haloes according to equation (\ref{eq:move}). This new catalogue can then be interpreted as a catalogue of haloes in the target cosmology at redshift $z'$, complete with new halo properties.
  \item If desired, reconstitute the particles in haloes using the method detailed in Section \ref{sec:reconstitution}.
\end{enumerate}

\section{Simulations}
\label{sec:simulations}

We illustrate our method using a matched set of simulations and the
halo catalogues generated from them. The simulation parameters are
given in Table \ref{tab:simulations}. The `target' simulation \lcdm is
a WMAP1 style cosmology (\citealt{Spergel2003}) run with the same
transfer function as that of the Millennium Simulation
(\citealt{Springel2005}) which was generated using \texttt{CMBFAST}
(\citealt{Seljak1996}). The `original' simulation \tcdm is a flat matter-only 
simulation run with a {\sc defw} transfer function
(\citealt{Davis1985}) tuned to have a similar spectral shape to that
of the Millennium Simulation. \tcdm models were popular in the past as
a way of enabling flat matter only models to fit clustering data from
contemporary galaxy surveys (\eg the APM survey:
\citealt{Maddox1990}) whose spectral shape appeared to require
a sub-critical mass density. The \tcdm model of
\cite{White1995} dealt with this by introducing extra relativistic
species, thus changing the epoch
of matter radiation equality
without lowering the mass density.

Initial
conditions were generated for each simulation by perturbing particle
positions from an initial glass configuration of $512^3$ particles
using the \texttt{N-GenIC} code at an initial redshift of
$z_{\mrm i}=199$. The simulations themselves were run using the cosmological
$N$-body code \texttt{Gadget-2} of \cite{Gadget2}.
Performing direct test simulations allows us to use the same
phases for the Fourier modes in the target and original simulations,
so that the approximate and exact target halo fields can be
compared visually, and not just at the level of power spectra.
This also allows us to analyse the results of the rescaling
without the added complication of cosmic variance.

\begin{table*}
\caption{Cosmological parameters for our simulations. As a `target' we
  use a \lcdm model with a WMAP1 type cosmology and as an `original'
  model we simulate a matter only model with a {\sc defw}
  (\citealt{Davis1985}) spectrum with a similar spectral shape
  ($\Gamma=0.21$) to that of the \lcdm model but that lacks a BAO
  feature. Each simulation ran with $512^3$ particles, gravitational
  forces were softened at $20\kpc$ and initial conditions generated
  using \texttt{N-GenIC} on an initial glass load at a starting
  redshift $z_{\mrm i}=199$.}  \centering
\begin{tabular}{c c c c c c c c c c}
\hline\hline 
Simulation & $L$ & $\Omega_{\mrm{m}}$ & $\Omega_{\Lambda}$ & $\Omega_{\mrm{b}}$ & $h$ & $\sigma_8$ & $n_{\mrm{s}}$ & $\Gamma$ \\ [0.5ex] 
\hline
\lcdm & $780\Mpc$ & 0.25 & 0.75 & 0.045 & 0.73 & 0.9 & 1 & - \\
\tcdm & $500\Mpc$ & 1 & 0 & - & 0.5 & 0.8 & 1 & 0.21 \\
\hline
\end{tabular}
\label{tab:simulations}
\end{table*}

\begin{table*}
\caption{Best fit scaling parameters for scaling between our original \tcdm model and our target \lcdm model.}
\centering
\begin{tabular}{c c c c c c c c c c c c}
\hline\hline 
Original & Target & $z$ & $z'$ & $s$ & $M_1$ & $M_2$ & $s_{\mrm m}$ & $k_{\mrm{nl}}$ & $b_{\mrm {eff}}$ \\[0.5ex] 
\hline
\tcdm & \lcdm & 0.22 & 0 & 1.56 & $\sform{2.58}{13}\Msun$ & $\sform{3.41}{15}\Msun$ & 0.95 & $0.15\iMpc$ & 1.59 \\
\hline
\end{tabular}
\label{tab:scaling}
\end{table*}

Our procedure for generating the simulations was as follows: run the
original simulation to $z=0$ in a box of size $L$, compile a halo
catalogue and then use the mass range in this halo catalogue to
compute the best scaling parameters ($s$, $z$) by minimising equation
(\ref{eq:minimise}). We then re-ran the original simulation to 
redshift $z$ because this used comparatively little computational resources.
However, in practice one would interpolate particle positions between 
simulation outputs around redshift $z$ if one was interested in particles,
or constrain the scaling redshift to be one of a set of $z$ (close to the best fit)
for which one already had an output. This would be necessary in the case of
halo catalogues because it is not obvious how to interpolate haloes
between catalogues due to mergers. For the purpose of comparisons we also ran a simulation of the target
cosmology to $z'=0$ in a box of size $L'=sL$ and compiled a halo catalogue. In
doing this step we chose the same random numbers for the mode phases and
amplitudes for the realization of the displacement fields to ensure
that structures appear in the same point in both simulations, despite
the different box sizes. This allowed for direct comparisons between the
simulation particle distributions and halo catalogues that are
affected only by the different background cosmologies rather than by
cosmic variance.

We compile halo catalogues using the public
FoF code
\texttt{www-hpcc.astro.washington.edu/tools/fof.html} with a linking
length of $0.2$ times the mean inter-particle separation in the
simulation. We catalogue only haloes that contain $\ge100$ particles
and we define halo centres to be the centre of mass of all
contributing halo particles.

For our simulations the best-fit scaling parameters are given in Table
\ref{tab:scaling}.  Fig. \ref{fig:mf} shows the effect of each stage of the scaling
on the halo mass functions; the theory of \cite{Sheth1999}
is shown in the top panel, whereas the effect on
the measured mass functions is shown in the bottom panel. The scaling
makes the theoretical predictions for the mass functions
agree to within 1\%, but this agreement is less perfect for
the measured mass functions, which display discrepancies of up to
10\%. This discrepancy can be traced back to
the fact that the fitting formula for the mass functions of
\cite{Sheth1999} are only accurate to 20\% and that the mass function
is only `nearly' universal (\citealt{Tinker2008};
\citealt{Lukic2007}). A similar level of disagreement in the measured mass
function was found in AW10 (their Fig. 7) in converting between WMAP1 and WMAP3
cosmologies.

Throughout this work we measure power spectra by creating the density
field on a mesh by assigning particles to mesh cells with a nearest
grid point mass assignment scheme and then taking the Fourier
transform of this density field. 
The effect of our
binning in cubic cells is corrected for by deconvolving the final
Fourier transform with the normalized transform of a cubic cell, \beq
J(k)=\sinc{(\theta_x)}\, \sinc{(\theta_y)}\, \sinc{(\theta_z)}\ , \eeq
where $\theta_i=k_i L/2 m$ and $m$ is the number of mesh cells used
for the density field. The power is thus adjusted according
to
\begin{equation}
\Delta^2(k)\rightarrow \frac{\Delta^2(k)}{J^2(k)}\ .
\end{equation}
We compute power spectra of both haloes and particles. In each case we
assume that these discrete objects randomly sample the (biased in the
case of haloes) mass field so we subtract shot noise from the power in
each case. This implies
\begin{equation}
\Delta^2(k)\rightarrow \Delta^2(k)-4\pi\left(\frac{k}{2\pi}\right)^3\frac{L^3}{N}\ ,
\end{equation}
where $N$ is the total number of particles in the simulation. This
correction is only important at small scales, where the shot noise
correction has been shown to be valid even for glass initial conditions (\citealt{Smith2003}).

\begin{figure*}
\centering
\subfloat{\includegraphics[width=58mm,angle=270,trim=2.4cm 4.5cm 1.1cm 6cm,clip,scale=1.35]{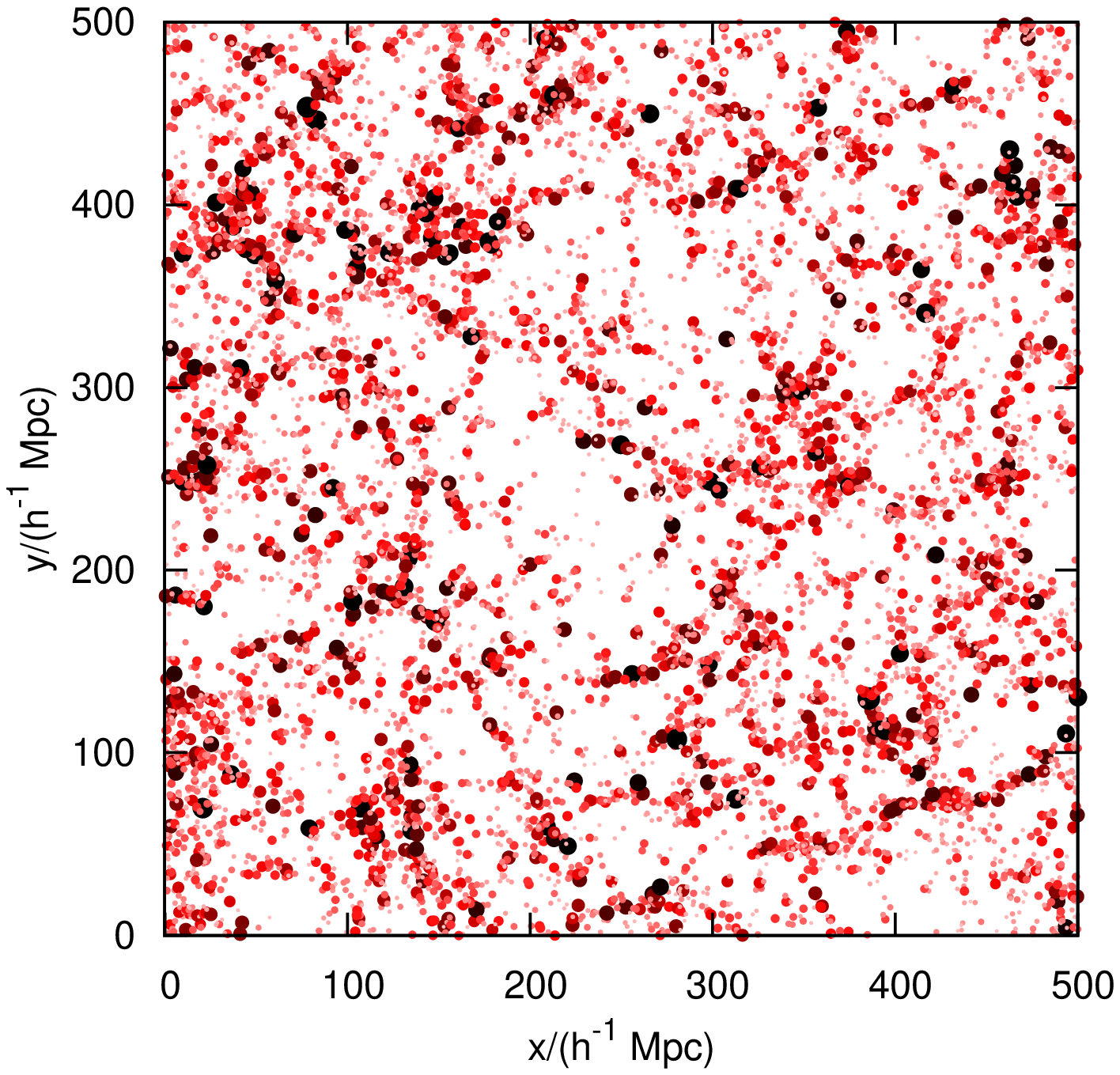}}\hspace{1cm}
\subfloat{\includegraphics[width=58mm,angle=270,trim=2.4cm 4.5cm 1.1cm 6cm,clip,scale=1.35]{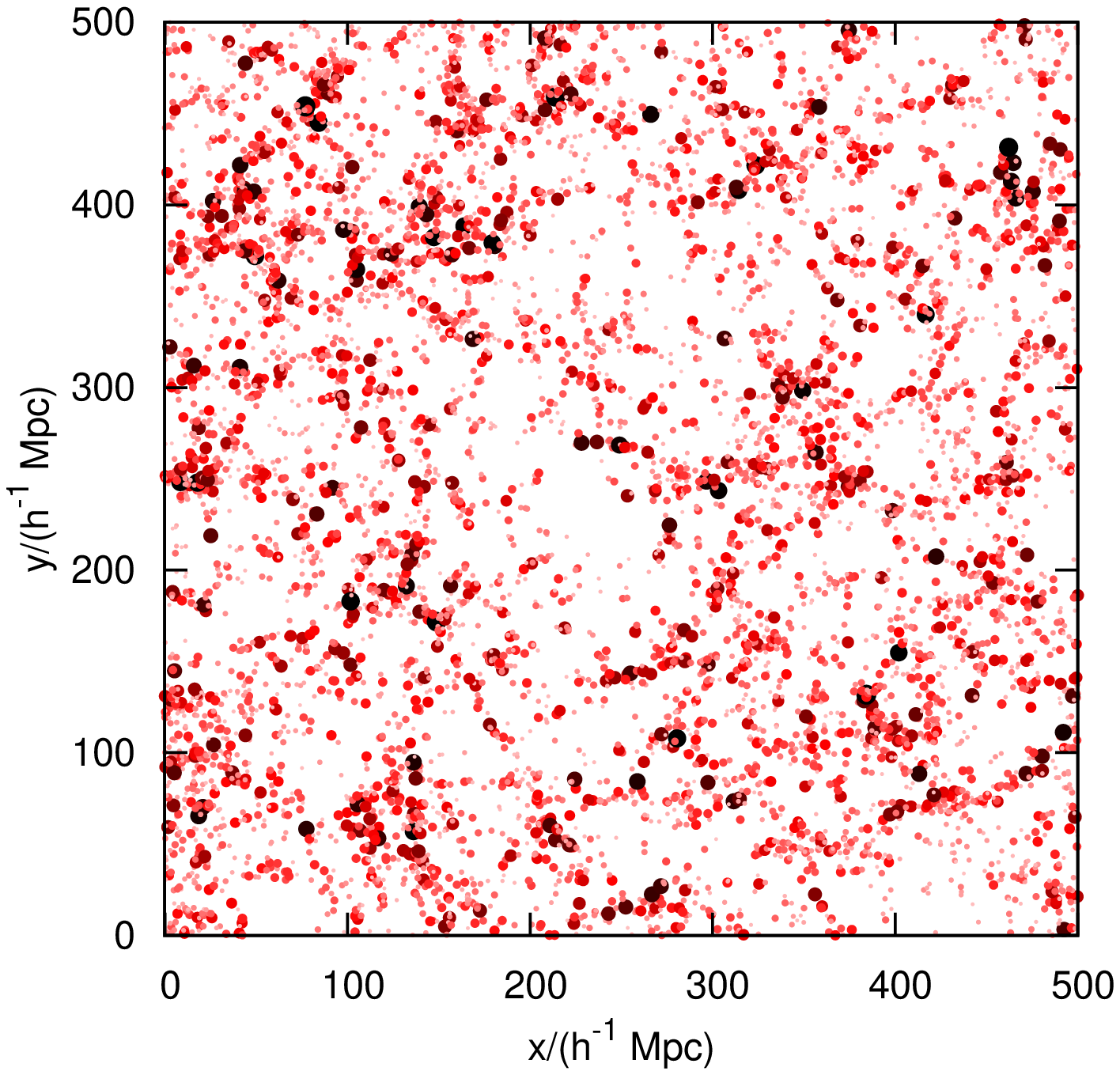}}\\
\subfloat{\includegraphics[width=58mm,angle=270,trim=2.4cm 4.5cm 1.1cm 6cm,clip,scale=1.35]{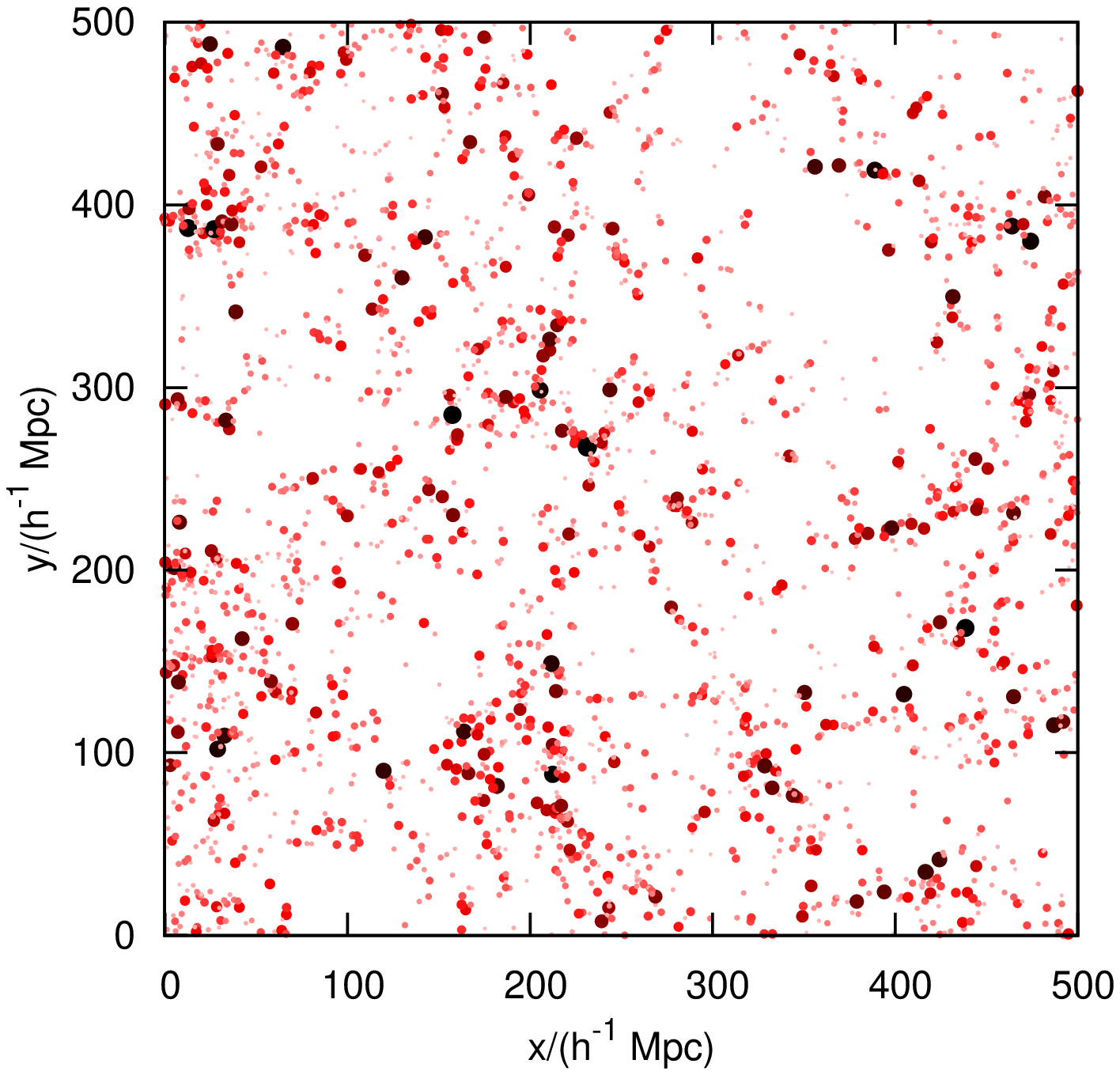}}\hspace{1cm}
\subfloat{\includegraphics[width=58mm,angle=270,trim=2.4cm 4.5cm 1.1cm 6cm,clip,scale=1.35]{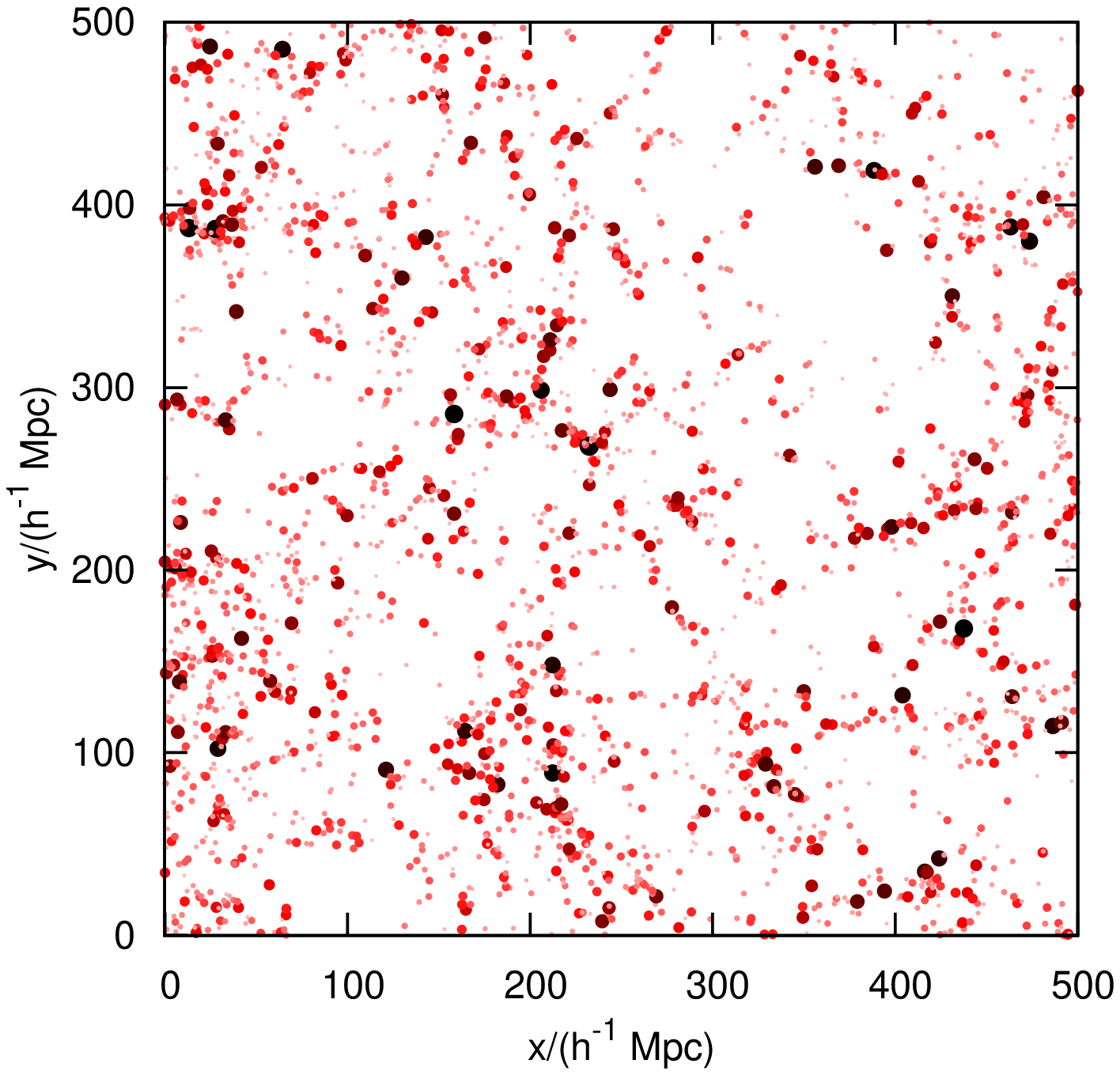}}\\
\caption{A pictorial summary of our results for the rescaling of a
  halo catalogue. The figures all show the halo distribution in
  $500\times 500\,(\Mpc)^2$ slices of thickness one tenth of the box size 
  ($50\Mpc$ for the upper two panels and $78\Mpc$ for the lower two) through
  different sections of our simulations. All panels show the haloes
  above $\sform{2.6}{13}\Msun$ with a point size and colour that depends on halo
  mass. The top left panel shows the halo distribution at redshift 0
  in the \tcdm simulation. The top right panel shows the distribution
  at redshift $z=0.22$ which is visibly less evolved. The bottom
  left panel shows the result of the full scaling algorithm; this
  mainly has the effect of a zoom owing to the scaling of box size and halo mass
  ($L\rightarrow 1.56L$ and $M\rightarrow 0.95M$), plus a shifting of haloes to reproduce the
  correct clustering according to the ZA. In fact, the ZA displacement
  is hard to detect by eye, but Fig. \ref{fig:halo_power} shows that
  it has a major impact on the halo power spectrum.  The bottom right
  panel shows the excellent agreement with the final halo distribution at
  redshift 0 in a directly constructed target \lcdm simulation, using
  the same phases as the rescaled \tcdm box.}
\label{fig:summary}
\end{figure*}

\section{Results from simulations}
\label{sec:results}

A visual summary of our rescaling method is given in
Fig. \ref{fig:summary}, where the distribution of haloes is shown at
each stage of the rescaling method. This illustrates the good agreement
between the distribution of haloes in the fully scaled
original halo catalogue and those in the target catalogue.
This comparison is facilitated by the fact that we are able to
use the same phases in the initial conditions for the two
simulations, so that any differences in appearance should
reflect only the treatment of nonlinear structure formation.

\begin{figure}
\begin{center}
\includegraphics[width=60mm,angle=270]{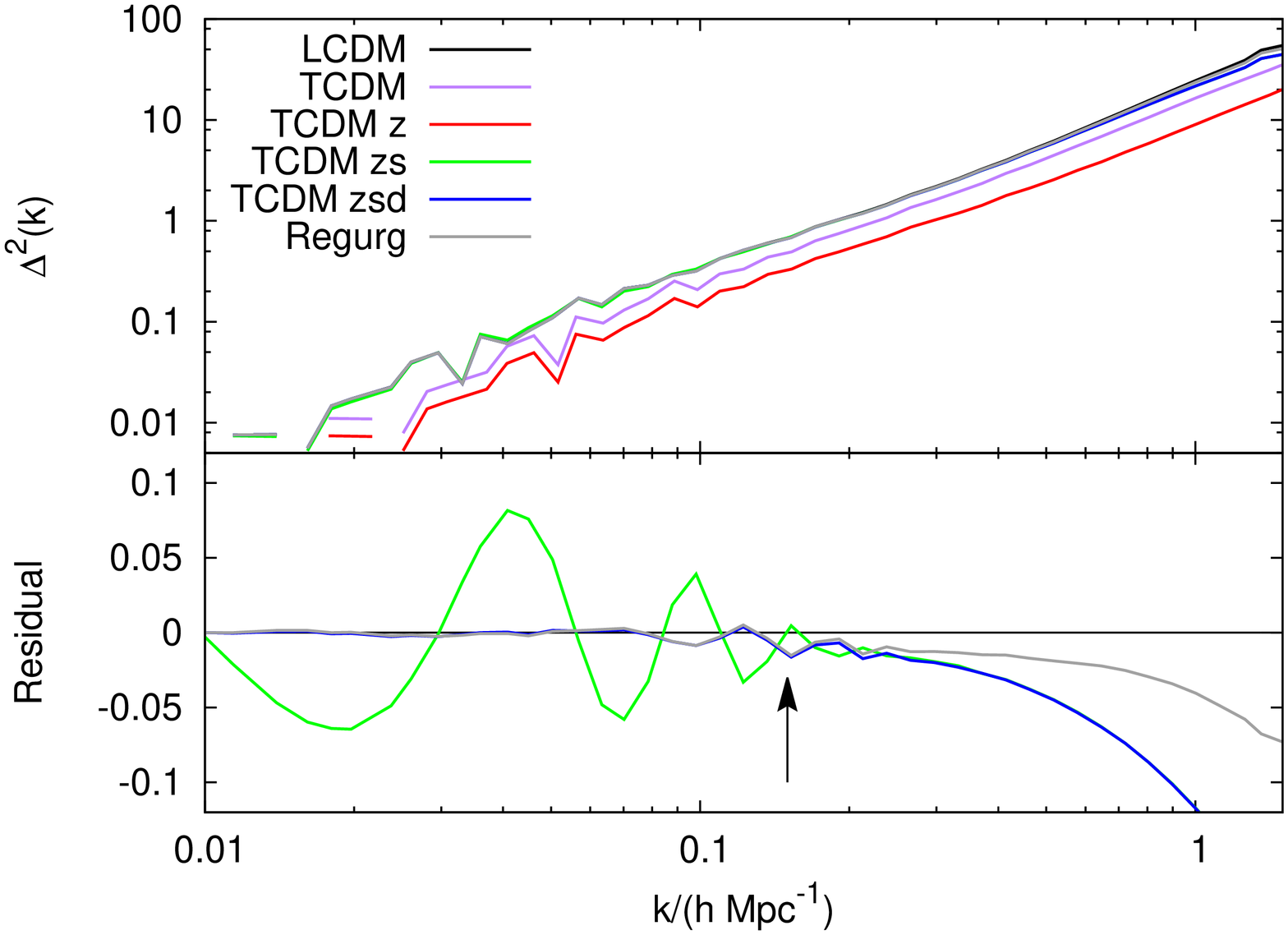}
\includegraphics[width=60mm,angle=270]{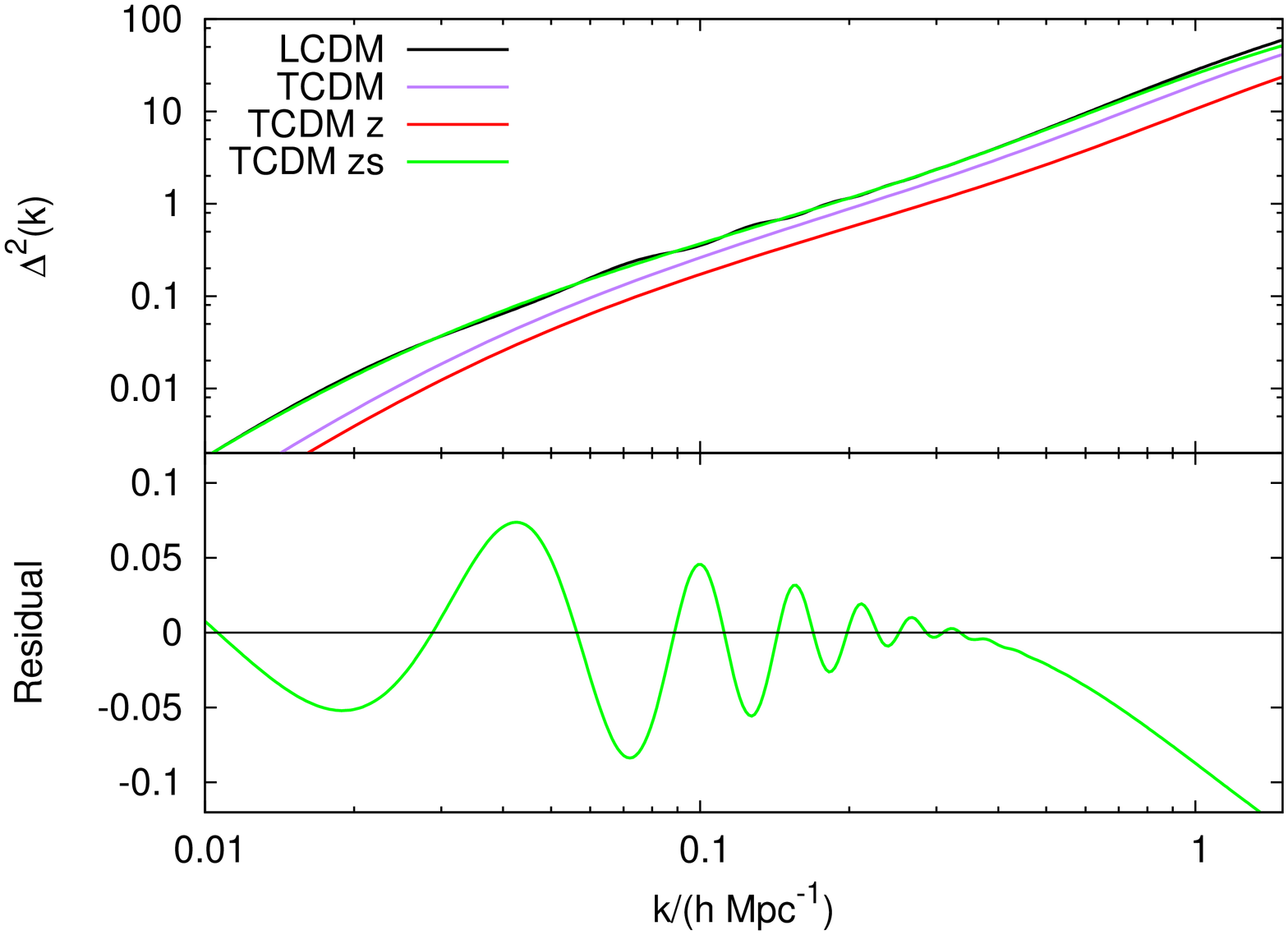}
\end{center}
\caption{The top panel shows the matter power spectra measured in the
  simulations at various stages of the method scaling between the
  \tcdm and \lcdm simulations. The black line is the target \lcdm
  power spectrum whereas the other lines show the various stages of
  the method, original \tcdm simulation (purple), scaling in redshift
  (red), scaling in redshift and size (green), full AW10 scaling
  including position modifications using the displacement fields
  (blue) and finally a power spectrum in which haloes have been
  removed from the scaled simulation and replaced with aspherical
  haloes of the correct concentrations for this cosmology
  (regurgitation, gray). One can see that adjusting particle
  positions using the ZA almost completely removes the residual BAO
  feature in the power spectrum, leaving the agreement between
  simulations at the level of $1\%$ up to the nonlinear scale (black
  arrow, equation \ref{eq:nl_scale}). Replacing haloes with those of
  the correct concentrations improves the match to the target
  simulation at small scales leaving the agreement at the level of
  $10\%$ up to $k=2\iMpc$. The lower panel shows the predicted
  differences for the full matter power spectrum given by a halo model
  calculation. The discrepancy seen at small scales in the halo model
  implies that this is due to halo internal structure and thus
  provides justification for modifying the halo internal properties.}
\label{fig:particle_power}
\end{figure}

As a first test of our method we aim to reproduce the AW10 results for
the power spectrum of the matter over-density field and these are shown
in Fig. \ref{fig:particle_power}. This is exactly the AW10 algorithm
except that we have regenerated the displacement fields from the
particle distribution directly. In these plots the full algorithm has
been applied to the particle distribution. The top panel shows the
measured power spectra at each stage of the scaling: One can
see that the BAO signal in the residual is completely removed
by modifying the particle positions and that the measured power
spectra agree at the 1\% level out to $k=0.15\iMpc$. Beyond this the
power spectra disagree at around the 20\% level, reflecting the fact
that the interior structure of the haloes has not been altered to
account for the change in background cosmology. We correct for
this using our reconstitution technique below. With this exception, it
is impressive that quite a broad shift in cosmological parameters (see
Table \ref{tab:simulations}) can be dealt with by the AW10
algorithm. This includes the generation of a BAO feature in the
particle distribution as well as the inclusion of vacuum energy -- even though
the results are based on the matter-only \tcdm simulation. This test is in very good
agreement with AW10 and provides a useful independent confirmation
of the accuracy of their algorithm. We have also compared the power spectrum 
obtained when using the original displacement field from the simulation (\ie the original
AW10 method), rather than reconstructed one, and we found negligible difference. This is good
given the scatter in the comparison of the displacement field see in Fig. \ref{fig:scatter}.
The bottom panel in Fig.
\ref{fig:particle_power} shows an analytical halo model prediction for the full
matter power spectrum, where we can see that the form of the rescaled
residual is very similar to that in the top panel. This motivates our
assertion that the remaining small-scale differences are due to the
treatment of halo internal structure.

\begin{figure}
\begin{center}
\includegraphics[width=60mm,angle=270]{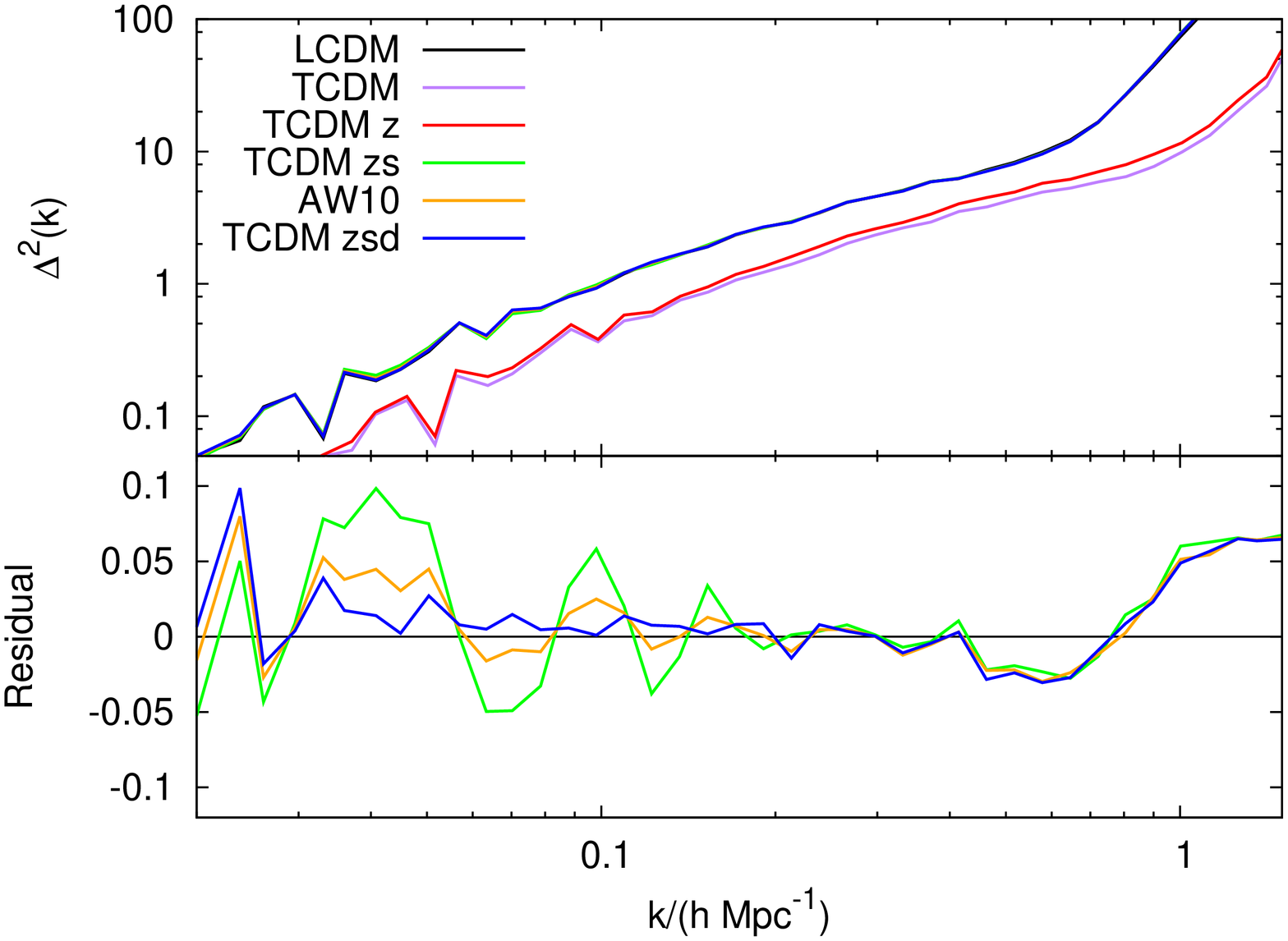}
\includegraphics[width=60mm,angle=270]{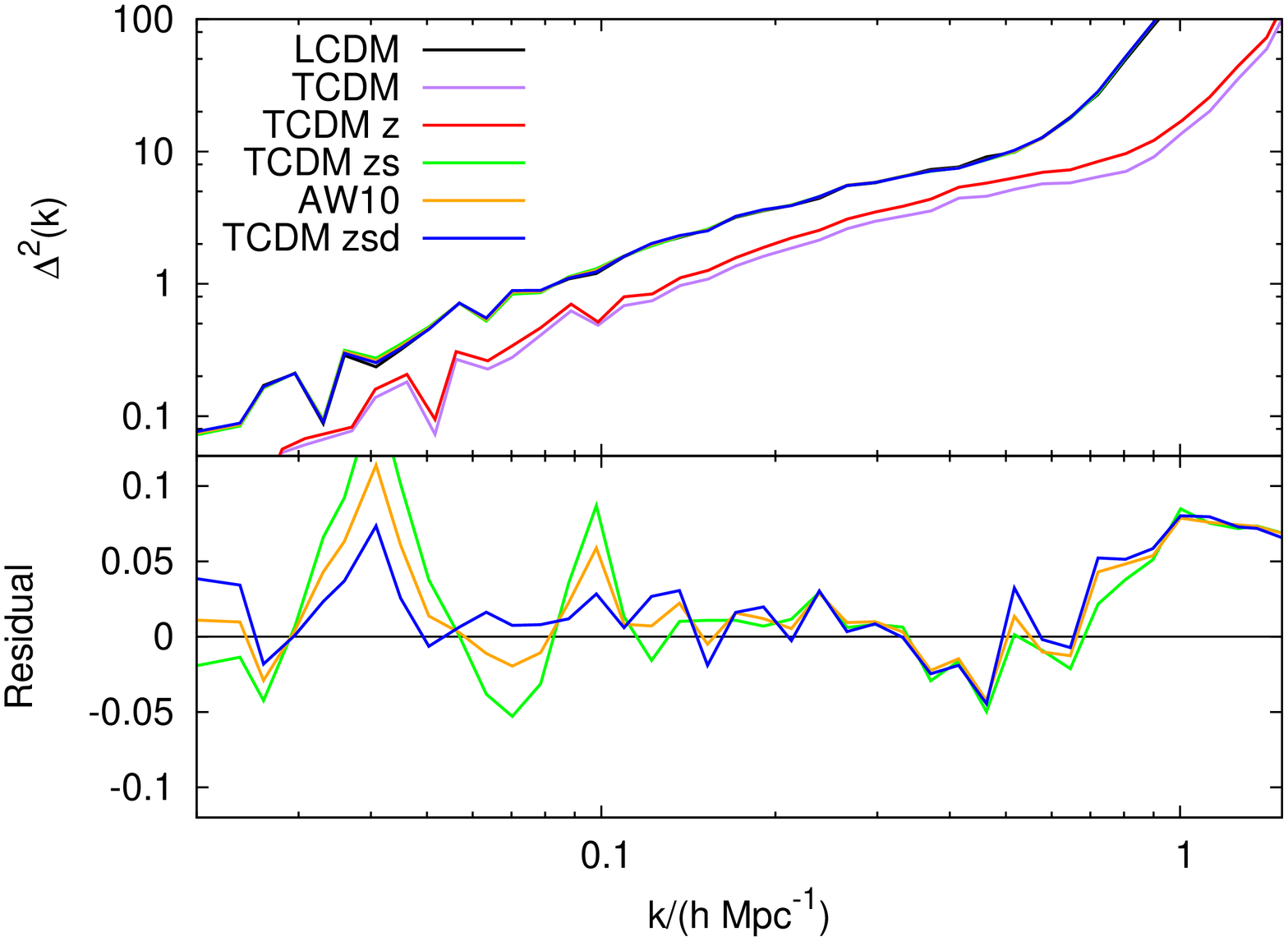}
\end{center}
\caption{The power spectrum of haloes at
  each stage of the rescaling procedure. We shows the target \lcdm
  spectrum of haloes (black) and the original halo catalogue at each
  stage of the scaling process, the original catalogue at $z=0$
  (purple), the redshift scaled \tcdm (red), \tcdm with both box size
  relabelled and redshift changed (green) and finally the result of
  also then modifying the halo positions according to the ZA. We
  do this in two distinct ways: applying the same differential
  displacement field to all haloes (AW10; orange), and giving different
  haloes a biased version of this displacement according to mass (blue)
  The upper panel shows the number weighted
  spectrum of haloes above $\sform{2.6}{13}\Msun$ while the lower panel shows that of
  haloes above $\sform{5}{13}\Msun$.
  In both cases, we can see that the universal displacement
  leaves a residual that reflects the BAO signal, whereas the mass-dependent
  displacement removes this problem entirely, leaving 
  agreement in both spectra at the level of $5\%$ or better except at the largest
  scales shown where the match is degraded slightly.}
\label{fig:halo_power}
\end{figure}

A more demanding test of rescaling is to ask if the method can
reproduce the desired clustering of haloes.
The results of our method of directly scaling a halo
catalogue are shown in Fig. \ref{fig:halo_power} as the number
weighted power spectra of haloes above $\sform{2.6}{13}\Msun$ in the
upper panel and the number weighted spectra of those above $\sform{5}{13}\Msun$ 
in the lower panel at each stage of the scaling process.
The displacement field required to
move haloes around according to the ZA has been generated entirely
from the halo distribution using the method described in the text.
Without this displacement, the power spectra are clearly in
error, with a residual that reflects the BAO signal.
This error is reduced when we apply the differential displacement
field to the haloes, but it is not eliminated. However, if the
displacement applied to each halo is scaled according to the
mass-dependent bias, $b(M)$, this problem is cured.
This confirms the need to apply a
mass-dependent differential displacement to haloes, an aspect
which is absent in the original AW10 algorithm.
As a further test of this point, we explicitly calculated the
mass-dependent bias to see if this was better recovered with our
method. But in practice the results were noisy (few haloes) and we are
looking for small shifts in a bias that is already well matched,
however, we believe the improved match in Fig. 9 is difficult to attribute to
anything other than an improved match in bias.
At the largest scales shown the rescaling method seems 
to degrade the match slightly and we are unsure why this is. However, we
see the same effect when we tested the method on smaller volume simulations
at the largest scales probed by \emph{those} simulations, scales that the method
shown in Fig. \ref{fig:halo_power} corrects well, so this is plausibly to
do with resolution on scales of order the box size.

\begin{figure}
\begin{center}
\includegraphics[width=60mm,angle=270]{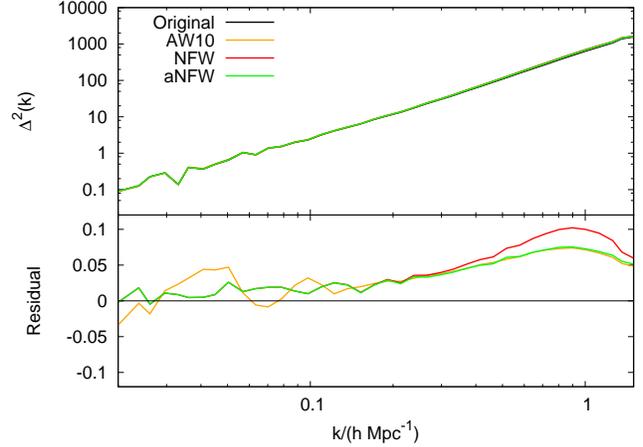}
\end{center}
\caption{The power spectra of only particles residing in haloes. We
  show spectra of the particles in the original haloes in the target
  simulation (black) together with particles in haloes reconstituted from the
  original catalogue that are spherical (red), aspherical (green). The orange curve shows
  the result of using the original AW10 method and aspherical haloes. Haloes are dressed
  with NFW profiles with concentration relations described in the
  text. There is a clear improvement gained by using a bias dependent displacement field
  and by using aspherical haloes over spherical haloes. The maximum error here does not exceed
  $6\%$.}
\label{fig:recon_halo_power}
\end{figure}

The final part of our investigations involves reconstituting the
particles in haloes using only the halo catalogues. In order to do
this we compare the power spectrum of only the particles in haloes
reconstituted from the \emph{scaled} \tcdm halo catalogue to the power spectrum of
particles in the haloes in the \lcdm simulation. This
is shown in Fig. \ref{fig:recon_halo_power}, which again displays
good agreement (the spectra agree to 5\% across the range
of scales shown) by using the full scaling algorithm. 
Clearly there is an improvement on small scales gained by using
aspherical haloes over spherical ones.

Finally we look at our regurgitation method in which, after the
original AW10 scaling method has been applied to particle data, the
haloes are located with FoF, removed and then replaced by reconstituted haloes with corrected
mass-concentration relations. The results of this were shown above in the
form of the power spectrum in
Fig. \ref{fig:particle_power}, where we see that the agreement between
the original and target cosmologies is much improved by this method at scales
above $k=0.1\iMpc$ due to our modifications of the haloes'
internal structure. Thus the final fully rescaled power spectra agrees
at a sub percent level to $k=0.1\iMpc$ and to a 5\% level out to
$k=1\iMpc$. Here there is a clear improvement over the original AW10
algorithm, gained by manipulating the properties of
individual haloes.

\section{Conclusions}
\label{sec:conclusions}

In this paper we have demonstrated that the rescaling method of
\cite{Angulo2010} may be modified so as to apply directly to halo
catalogues.
We made
an AW10 rescaling of length, mass, and
redshift as well as using the halo positions themselves to compute the
displacement fields (by debiasing the halo over-density fields), in order
to correct the linear clustering in the simulation
using the Zel'dovich approximation. 
This method enables rapid scaling of a halo catalogue to a
different cosmology, and is entirely self-contained,
being based only on the halo catalogue. We note that this provides
a dramatic increase in speed when using the halo catalogue alone. Computational
effort is only really used when reading the catalogue into memory and when computing
the Fourier Transforms for the displacement field correction. In our case the halo
catalogue was small, containing only $\sim 70,000$ haloes, and a Fourier mesh of only $75^3$
was all that was required to resolve the linear components of the displacement field. This resulted
in a total run time for the rescaling of only a few seconds on a standard desktop computer. This would increase for larger
volume halo catalogues because more mesh cells would be required to resolve the linear fields, and for
larger halo catalogues, but it is obviously many orders of magnitude faster than running a new simulation.

Working with haloes has the advantage of speed, but also
allows two improvements on the original AW10 method.
The first of these concerns the internal structure of haloes,
which depends on cosmology. This can be allowed for
by `reconstituting' the halo internal density distribution
using analytical profiles and scaling relations appropriate
for the target cosmology. If the catalogue of halo
particles is available, it is also possible to
`regurgitate', in which haloes are replaced with those
with the correct internal structure for the new cosmology.


The other issue applies on large scales. The AW10 method applies an additional
displacement in order to ensure that the large-scale linear 
clustering is as desired in the target cosmology. But applying
this extra displacement to all haloes, independent of their mass,
will not yield the correct mass-dependent bias, $b(M)$. 
We found that better results were obtained at the level of the power spectrum by scaling the
extra displacement in a mass-dependent way. Clearly this is a minor issue if the original and target cosmologies are
close to each other, but it may be important in spanning a large parameter space.

We have tested our method by rescaling a halo catalogue generated from
a matter-only \tcdm simulation into that of a more standard \lcdm
model. This represents a radical shift in cosmology, especially
considering that the initial simulation contains
no dark energy. At the
level of the particle distribution the matter power spectrum is
predicted correctly after the rescaling to the level of 1\% to
$k=0.2\iMpc$ and 10\% to $k=1\iMpc$. This is in excellent agreement
with the original AW10 results and provides independent confirmation of the 
accuracy of the scaling algorithm. For the haloes the
power spectra are noisier, but are still predicted correctly at the
level of 10\% up to $k=0.2\iMpc$ with no obvious biases.
We have also tested reconstitution of haloes, showing that
this method is
able to reproduce the power spectrum of particles in haloes at the
level of 2\% at $k=1\iMpc$ in the perfect case 
(no rescaling, simply comparing reconstituted halo particles to those originally used to compile
the catalogues) and 7\% in the case of
our reconstitution of the rescaled haloes. By replacing haloes in a scaled particle distribution
 via `regurgitation', a further improvement is gained over the
original AW10 algorithm, leaving
the agreement in matter power spectra essentially perfect at
scales below $k=0.1\iMpc$ and agreeing to within 5\% to $k=1\iMpc$. 
More demanding tests of the method are certainly possible; in future
work we aim to consider aspects beyond the power spectrum, for example
higher-order statistics or details of the differences in position
between rescaled and target haloes.

We have also provided a method for rescaling velocities, although
a detailed investigation of the operation of this method in
redshift space will be given elsewhere.
Another interesting line of investigation would be to
see if the displacement field can be better reproduced from the halo
distribution by looking at different reconstruction
techniques. Further work could also examine the effect of
different mass-concentration relations for the haloes or
different non-universal prescriptions for the mass function.
Even so, the current method already seems well suited for the application
of rapid generation of mock galaxy catalogues covering a wide
range of cosmologies.

\section*{Acknowledgements}

AJM acknowledges the support of an STFC studentship.

\label{lastpage}
\setlength{\bibhang}{2.0em}
\setlength\labelwidth{0.0em}
\bibliographystyle{mn2e}                      
\bibliography{japbib}
\end{document}